\documentclass[sigconf, nonacm]{acmart}
\AtBeginDocument{%
  }

\setcopyright{cc}
\copyrightyear{2025}
\acmYear{2025}
\acmDOI{}
\acmConference[EDBT '26]{Make sure to enter the correct
  conference title from your rights confirmation email}{24-27 March,
  2026}{Tampere (Finland)}
\acmISBN{}

\usepackage{colortbl} 
\usepackage{xcolor}   
\definecolor{lightgreen}{RGB}{198,239,206}
\definecolor{lightgray}{RGB}{217,217,217}
\usepackage{listings}
\definecolor{codegreen}{rgb}{0,0.6,0}
\definecolor{codegray}{rgb}{0.5,0.5,0.5}
\definecolor{codepurple}{rgb}{0.58,0,0.82}
\definecolor{backcolour}{rgb}{0.95,0.95,0.92}
\lstdefinestyle{mystyle}{
    commentstyle=\color{codegreen},
    numberstyle=\tiny\color{codegray},
    stringstyle=\color{codepurple},
    basicstyle=\ttfamily\footnotesize,
    breakatwhitespace=false,         
    breaklines=true,                 
    captionpos=b,                    
    keepspaces=true,                 
    numbers=left,                    
    numbersep=4pt,                  
    showspaces=false,                
    showstringspaces=false,
    showtabs=false,                  
    tabsize=1
}

\usepackage{subcaption}
\begin{document}

\title{MobilityDuck: Mobility Data Management with DuckDB}

\author{Nhu Ngoc Hoang$^*$}
\affiliation{
    \institution{Université Libre de Bruxelles}
    \city{Brussels}
    \country{Belgium}
}
\email{nhu.hoang@ulb.be}

\author{Ngoc Hoa Pham$^*$}
\affiliation{
    \institution{Université Libre de Bruxelles}
    \city{Brussels}
    \country{Belgium}
}
\email{ngoc.pham@ulb.be}

\author{Viet Phuong Hoang$^*$}
\affiliation{
    \institution{Université Libre de Bruxelles}
    \city{Brussels}
    \country{Belgium}
}
\email{viet.hoang@ulb.be}

\author{Esteban Zimányi$^*$}
\affiliation{
    \institution{Université Libre de Bruxelles}
    \city{Brussels}
    \country{Belgium}
}
\email{esteban.zimanyi@ulb.be}

\renewcommand{\shortauthors}{Nhu Ngoc Hoang, Ngoc Hoa Pham, Viet Phuong Hoang, and Esteban Zimányi}


\begin{abstract}
  The analytics of spatiotemporal data is increasingly important for mobility analytics. Despite extensive research on moving object databases (MODs), few systems are ready on production or lightweight enough for analytics. MobilityDB is a notable system that extends PostgreSQL with spatiotemporal data, but it inherits complexity of the architecture as well. In this paper, we present MobilityDuck, a DuckDB extension that integrates the MEOS library to provide support spatiotemporal and other temporal data types in DuckDB. MobilityDuck leverages DuckDB's lightweight, columnar, in-memory executable properties to deliver efficient analytics. To the best of our knowledge, no existing in-memory or embedded analytical system offers native spatiotemporal types and continuous trajectory operators as MobilityDuck does. We evaluate MobilityDuck using the BerlinMOD-Hanoi benchmark dataset and compare its performance to MobilityDB. Our results show that MobilityDuck preserves the expressiveness of spatiotemporal queries while benefiting from DuckDB’s in-memory, columnar architecture.
\end{abstract}





\maketitle
\noindent $^*$All authors contributed equally to this work.

\section{Introduction}
The rapid growth of spatiotemporal data has created new opportunities for mobility analytics, where discovering patterns and trends in object trajectories plays a central role in applications such as urban planning, intelligent transportation systems, and mobility-as-a-service platforms.

 Despite an extensive body of research in moving object databases (MODs), and the emergence of systems like MobilityDB, mainstream adoption is still limited by architectural complexity, setup overhead, and integration challenges in modern analytics pipelines. However, MobilityDB inherits PostgreSQL’s complexity, which limits its efficiency for lightweight querying, embedded deployment, and exploratory data science workflows where ease of use and speed of integration are paramount.

 At the same time, DuckDB has rapidly emerged as a modern analytical database, designed to be lightweight, embeddable, and highly optimized for in-memory, columnar query execution. Nevertheless, DuckDB currently lacks first-class support for spatiotemporal data types and operators.

This paper introduces MobilityDuck, the first DuckDB extension to support spatiotemporal and temporal data types . By combining DuckDB’s in-memory, vectorized execution model with MEOS’s mature spatiotemporal algebra, MobilityDuck brings the expressiveness of moving object databases into a lightweight analytical engine. We also adapt the BerlinMOD benchmark to the Hanoi urban environment, producing BerlinMOD-Hanoi, a reproducible dataset and query workload for diverse mobility analytics. Our experimental evaluation shows that MobilityDuck maintains query expressiveness while delivering significant performance improvements on most benchmark tasks.

\section{Background and Related Work}
\subsection{Spatiotemporal Data Management}
Research on spatiotemporal data management has a long history in both the database and GIS communities. Early efforts studied spatial and temporal aspects separately, combining them later through extensions of existing database systems. Some proposals extended spatial databases with temporal versioning (e.g.,\cite{newell1992temporal}),  while others extended temporal databases with spatial types and attributes. 
Systems such as TGRASS \cite{gebbert2017grass} integrated time with 2D and 3D spatial fields to enable space-time analysis, organizing data as snapshots into space-time fields. A comprehensive review of these early models can be found in \cite{pelekis2004literature}.

Beyond discrete temporal tagging, another research direction aimed to model continuously evolving objects. Constraint databases \cite{grumbach1998spatio} provided a theoretical foundation for representing spatiotemporal entities as sets of points defined by constraints. The DEDALE system \cite{grumbach1997dedale} implemented this model, allowing relational algebra to be performed efficiently over 3D (2D space + 1D time) objects. 

A parallel and more practical line of work followed the \emph{abstract data type (ADT)} approach, where spatiotemporal types and operations are implemented natively inside extensible database systems. This approach led to mature prototypes such as SECONDO \cite{guting2005secondo}, which defines an extensible algebra for moving objects, including types such as \texttt{mpoint} and indexes such as RTree and TBTree. 

For large-scale and distributed settings, systems such as Parallel SECONDO \cite{lu2013parallel}, Distributed SECONDO \cite{nidzwetzki2017distributed}, Geomesa \cite{hughes2015geomesa}, ST-Hadoop \cite{alarabi2018st}, TrajSpark \cite{hagedorn2017stark}, GeoFlink \cite{shaikh2020geoflink} have explored the integration of spatiotemporal data management into Hadoop, Spark, and Flink. These systems provide global indexes and parallel operators to efficiently distribute trajectory data and queries across clusters.

In addition, several ISO \cite{iso19141} and OGC \cite{ogc2010, ogc2013, ogc2014, ogc2016, ogc2018, ogc2019} standards have been proposed for representing and exchanging moving feature data. More recently, the spatial data ecosystem has expanded these efforts through open columnar. The OGC GeoParquet 1.1.0 specification \cite{holmes2024geoparquet} extends the Apache Parquet format to support geometry columns and spatial metadata, while the forthcoming Apache Parquet release introduces native geometry and geography support \cite{dem2025parquet}. 
Similarly, the GeoArrow 0.1.0 specification \cite{dunnington2024geoarrow} defines Arrow extension types and memory layouts for geometries compatible with analytical systems such as DuckDB, Polars, and cuDF, improving interoperability between storage and in-memory analytics. 
These initiatives reflect a convergence between traditional geospatial standards and modern analytical ecosystems.

Among the many research prototypes, MobilityDB \cite{Zimanyi2019} has emerged as the most complete  open-source implementation of a moving object database \cite{MobilityDataScience2025}. It extends PostgreSQL and PostGIS with \emph{temporal types} and \emph{spatiotemporal operators}, building on the MEOS (Mobility Engine Open Source) library. It supports moving points (e.g., vehicle trajectories), temporal spans, and temporal aggregates. MobilityDB has become a reference implementation for managing mobility data, but inherits PostgreSQL’s overhead in query execution and storage management. However, its performance remains limited by PostgreSQL’s general-purpose query engine and storage layer.

Motivated by the need for faster analytical processing and simpler deployment, there has been growing interest in in-memory and memory-efficient architectures for spatiotemporal data. S4STRD presents a scalable in-memory storage system for real-time trajectory data, keeping recent updates in RAM and using NoSQL backends for persistence \cite{pham2015s4strd}. SharkDB \cite{wang2014sharkdb} is an in-memory, column-oriented trajectory storage system that partitions trajectories into time-based frames, allowing efficient compression, memory throughput, and parallel processing across cores. In a complementary direction, Richly et al. propose optimized spatio-temporal data structures for in-memory columnar databases, adapting memory layouts, compression, and tiering to trajectory workloads \cite{richly2021memory}. These works illustrate the feasibility and challenges of in-memory spatiotemporal storage -particularly for reducing I/O overhead, but they emphasize storage and access optimizations rather than full query semantics.
In contrast, MobilityDuck embeds spatiotemporal types and operators directly into an analytical SQL engine, enabling expressive querying over moving object data within the DuckDB ecosystem. 

\subsection{The MEOS Library}
At the core of MobilityDB is the \textbf{MEOS} library \cite{Zimanyi2024}, a C library that implements temporal and spatiotemporal data types and functions independently of PostgreSQL. 

MEOS extends the ISO 19141:2008 \cite{iso19141} standard (Geographic information—Schema for moving features) for representing the change of non-spatial attributes of features. It also takes into account the fact that when collecting mobility data it is necessary to represent “temporal gaps”, that is, when for some period of time no observations were collected due, for instance, to signal loss.

MEOS is inspired by a similar library called GEOS (Geometry Engine, Open Source) — hence the name. A first version of the MEOS library written in C++ has been proposed by Krishna Chaitanya Bommakanti. However, due to the fact that MEOS codebase is actually a subset of MobilityDB codebase, which is written in C and in SQL, the current version of the library allows us to evolve both programming environments simultaneously.

MEOS supports generic temporal types (e.g., \texttt{tbool}, \texttt{tint}, \texttt{tfloat}, \texttt{ttext}) and spatiotemporal types (e.g., \texttt{tgeompoint}, \texttt{tgeogpoint}), together with indexing, synchronization, and aggregation operators.  This separation allows other systems, such as DuckDB in our case, to reuse MEOS without relying on PostgreSQL.  

\subsection{Benchmarks for Moving Object Databases}
Evaluating spatiotemporal DBMSs requires reproducible benchmarks. \textbf{BerlinMOD}~\cite{Dntgen2009} is the standard benchmark for moving object databases. It defines a synthetic mobility model, a trip generation based on an underlying road-network, and a set of queries measuring performance on indexing, joins, and aggregates.  
MobilityDB has been evaluated extensively using BerlinMOD, which makes it a natural baseline for our work.  

To adapt BerlinMOD to different geographic contexts, in this paper, we introduced \textbf{BerlinMOD-Hanoi} (see Section~\ref{sec:berlinmod-hanoi}), which applies the BerlinMOD benchmark using the Hanoi road network from OpenStreetMap data as base map.

\subsection{DuckDB and In-process Analytics}
DuckDB\footnote{\href{https://duckdb.org/}{https://duckdb.org}} is an open-source relational database management system developed by Mark Raasveldt and Hannes Mühleisen \cite{Raasveldt2019}. DuckDB is optimized for online analytical processing (OLAP) workloads, making it a suitable system for handling complex querying on large datasets \cite{raasveldt2020data}. The key features of DuckDB are as follows:

\begin{itemize}
    \item \textbf{Embeddability}: Unlike traditional database systems with large servers running as stand-alone processes, DuckDB is designed to be an embedded database system that runs completely within another host process.
    \item \textbf{Analytical}: While other embedded systems (e.g., SQLite) focus more on transactional (OLTP) workloads, DuckDB is geared towards efficiently executing analytical SQL queries.
    \item \textbf{High performance}: DuckDB employs a vectorized interpreted execution engine, which optimizes CPU cache usage and allows batch processing of data.
    \item \textbf{Integration with other tools}: DuckDB supports complex SQL queries and provides APIs for a wide range of programming languages, namely C++, Java, Python, Rust, Swift, among others. Existing popular interactive data analysis tools such as the \verb|dplyr| package in R or the \verb|pandas| library in Python can be used alongside DuckDB, which addresses the lack of support for query optimization and transactional storage in these tools.
\end{itemize}

Recent work has extended DuckDB with domain-specific extensions (e.g., for geospatial analytics via DuckDB Spatial Extension\cite{gabrielsson2023postgeese}, for machine learning via QuackML \cite{quackML2025}). However, there is no native support for spatiotemporal types and operators.  

\section{MobilityDuck: Architecture and Implementation}

\subsection{Design Goals}
Our primary goal with \textbf{MobilityDuck} is to enable spatiotemporal analytics within DuckDB by reusing the mature functionality of the MEOS library. The design is guided by the following principles:
\begin{itemize}
    \item \textbf{Lightweight integration:} MobilityDuck is implemented as a DuckDB extension, preserving DuckDB’s embedded deployment model.  
    \item \textbf{Reuse of MEOS:} Instead of reimplementing temporal types and operators, we wrap MEOS natively in C++, ensuring correctness and consistency with MobilityDB.
    \item \textbf{DuckDB compatibility:} All types and functions are exposed as DuckDB user-defined types (UDTs) and functions, allowing seamless integration with DuckDB’s SQL engine, storage manager, and vectorized execution model.  
\end{itemize}

\subsection{System Architecture}
MobilityDuck follows a simple and modular architecture that connects DuckDB with the MEOS library through a thin C++ extension layer. At query time, DuckDB executes SQL statements as usual, while the extension intercepts calls to spatiotemporal functions and forwards them to MEOS.

Conceptually, the system has three main layers:
\begin{itemize}
    \item \textbf{DuckDB core:} provides the SQL parser, planner, storage engine, and vectorized execution framework. 
    MobilityDuck registers its custom types and functions within this engine at load time.
    \item \textbf{MobilityDuck extension layer:} acts as the bridge between DuckDB and MEOS. It defines DuckDB user-defined types and functions (e.g., \texttt{tint}, \texttt{tfloat}, \texttt{span}) based on their corresponding MEOS structures.    
    \item \textbf{MEOS library:} provides the underlying temporal and spatial operators and data structures used by MobilityDB. 
\end{itemize}
This design ensures minimal overhead while maintaining full compatibility with existing DuckDB operations.

\subsection{Type System and Registration}
All types in \textbf{MobilityDuck} follow the same design as in \textbf{MobilityDB}, but require explicit registration in DuckDB. Internally, all MEOS types are represented using the native DuckDB type \texttt{BLOB}, allowing them to encode arbitrary binary objects while preserving type safety through the extension’s type system. 

For example, the bounding box type (\texttt{stbox}), which is composed of spatial and/or temporal dimensions, is implemented as follows:
\begin{lstlisting}[language=C++,numbers=none]
LogicalType StboxType::STBOX() {
  LogicalType type(LogicalTypeId::BLOB);
  type.SetAlias("STBOX");
  return type;
}
void StboxType::RegisterType(DatabaseInstance 
    &instance) {
  ExtensionUtil::RegisterType(instance, "STBOX", 
    STBOX());
}
\end{lstlisting}

Here, the underlying representation is a \texttt{BLOB}, while the alias ensures that queries can refer to the type as \texttt{stbox}, consistent with MobilityDB.

\textbf{Supported data types.} Currently, MobilityDuck exposes a subset of MEOS types as first-class DuckDB types. The coverage is summarized in Table~\ref{tab:types}: green cells indicate types already implemented in MobilityDuck, white cells are available in MobilityDB but not yet implemented, and gray cells are not applicable.

\begin{table}[h]
\caption{Template types supported in \textbf{MobilityDB} and in \textbf{MobilityDuck}.
\label{typetable}
Green: supported in MobilityDuck and in MobilityDB, White: in MobilityDB only, Gray: not applicable.}
\centering
\renewcommand{\arraystretch}{1.2}
\setlength{\tabcolsep}{2pt}
\begin{tabular}{|l|c|c|c|c|}
\cline{2-5}
\multicolumn{1}{c|}{} & \multicolumn{4}{c|}{\textbf{Template types}} \\
\hline
\textbf{Base types} & \textbf{set} & \textbf{span} & \textbf{spanset} & \textbf{temporal} \\
\hline
\textbf{bool}        & \cellcolor{lightgray} & \cellcolor{lightgray} & \cellcolor{lightgray} & \cellcolor{lightgreen} tbool \\
\hline
\textbf{text} & \cellcolor{lightgreen} textset & \cellcolor{lightgray} & \cellcolor{lightgray} & \cellcolor{lightgreen} ttext \\
\hline
\textbf{integer}     & \cellcolor{lightgreen} intset & \cellcolor{lightgreen} intspan & \cellcolor{lightgreen} intspanset & \cellcolor{lightgreen} tint \\
\hline
\textbf{bigint} & \cellcolor{lightgreen} bigintset & \cellcolor{lightgreen} bigintspan & \cellcolor{lightgreen} bigintspanset & \cellcolor{lightgray} \\
\hline
\textbf{float} & \cellcolor{lightgreen} floatset & \cellcolor{lightgreen} floatspan & \cellcolor{lightgreen} floatspanset & \cellcolor{lightgreen} tfloat \\
\hline
\textbf{date} & \cellcolor{lightgreen} dateset & \cellcolor{lightgreen} datespan & \cellcolor{lightgreen} datespanset & \cellcolor{lightgray} \\
\hline
\textbf{timestamptz} & \cellcolor{lightgreen} tstzset & \cellcolor{lightgreen} tstzspan & \cellcolor{lightgreen} tstzspanset & \cellcolor{lightgray} \\
\hline
\textbf{geometry}    & \cellcolor{lightgreen} geomset & \cellcolor{lightgray} & \cellcolor{lightgray} &
  \cellcolor{lightgreen} \begin{tabular}{@{}c@{}} tgeompoint \\ tgeometry \end{tabular} \\
\hline
\textbf{geography}   & geogset & \cellcolor{lightgray} & \cellcolor{lightgray} &
  \begin{tabular}{@{}c@{}} tgeogpoint \\ tgeography \end{tabular} \\
\hline
\textbf{pose} & poseset & \cellcolor{lightgray} & \cellcolor{lightgray} &  tpose \\
\hline
\textbf{npoint} &  npointset & \cellcolor{lightgray} & \cellcolor{lightgray} & tnpoint \\
\hline
\textbf{cbuffer} & cbufferset & \cellcolor{lightgray} & \cellcolor{lightgray} & tcbuffer \\\hline
\end{tabular}
\label{tab:types}
\end{table}

\subsection{Registration of Functions and Operators}
MobilityDuck exposes functionality through three categories of functions.

\vspace{1mm}\noindent
\textbf{Cast functions\hspace{1mm}} These implement explicit conversions between MobilityDuck types. A custom cast functions must be defined with a specific signature, for example with \texttt{Tbox}: 
\begin{lstlisting}[language=C++,numbers=none]
bool TboxFunctions::Tbox_in(Vector &source, Vector &result, idx_t count, CastParameters &parameters);
\end{lstlisting}
    Once defined, the cast function is registered in DuckDB as follows:
\begin{lstlisting}[language=C++,numbers=none]
void TboxType::RegisterCastFunctions(DatabaseInstance 
    &instance) {
  ExtensionUtil::RegisterCastFunction(
    instance,
    LogicalType::VARCHAR,   // input
    TBOX(),                 // output
    TboxFunctions::Tbox_in  // function
  );
}
\end{lstlisting}

\vspace{1mm}\noindent
\textbf{Scalar functions\hspace{1mm}} Other functions are defined as scalar functions, which have signatures different from cast functions.  For example, the following functions operate on \texttt{Set} types:
\begin{lstlisting}[language=C++,numbers=none]
static void Value_to_set(DataChunk &args, 
  ExpressionState &state, Vector &result);
static void Intset_to_floatset(DataChunk &args, 
  ExpressionState &state, Vector &result);
static void Floatset_to_intset(DataChunk &args, 
  ExpressionState &state, Vector &result);
static void Dateset_to_tstzset(DataChunk &args, ExpressionState &state, Vector &result);
static void Tstzset_to_dateset(DataChunk &args, 
  ExpressionState &state, Vector &result);    
static void Set_mem_size(DataChunk &args, 
  ExpressionState &state, Vector &result);
\end{lstlisting}    
    All scalar functions then need to be registered with DuckDB. Below is an example on how to register \texttt{shiftScale()} function with DuckDB: 
\begin{lstlisting}[language=C++]
void SetTypes::RegisterScalarFunctions(DatabaseInstance &db) {
  ExtensionUtil::RegisterFunction(
    db,
    ScalarFunction("shiftScale", 
    {SetTypes::intset(), LogicalType::INTEGER,
    LogicalType::INTEGER}, SetTypes::intset(), 
    SetFunctions::Numset_shift_scale)
  );
}
\end{lstlisting}

\vspace{1mm}\noindent
\textbf{Operators\hspace{1mm}} Unlike PostgreSQL, DuckDB does not provide a separate \texttt{CREATE OPERATOR} statement. 
    Instead, operators can be defined as binary scalar functions, using the operator symbol as the function name. The left operand becomes the first parameter, and the right operand becomes the second. When registering operators with DuckDB, the operator symbol is used as the function name:
\begin{lstlisting}[language=C++,numbers=none]
ExtensionUtil::RegisterFunction(
  instance,
  ScalarFunction(
    "&&", // overlaps
    {TGEOMPOINT(), StboxType::STBOX()},
    LogicalType::BOOLEAN,
    TgeompointFunctions::Temporal_overlaps_tgeompoint_stbox
  )
);
\end{lstlisting}

\subsection{Sample Queries}
This section introduces a number of sample queries utilizing different mobility types and functions available in MobilityDuck.
\begin{itemize}
    \item Return the time interval of a temporal type:
\begin{lstlisting}[language=SQL,numbers=none]
SELECT duration('{1@2025-01-01, 2@2025-01-02, 
  1@2025-01-03}'::TINT, true);
-- 2 days
\end{lstlisting}
    \item Shift and scale a \texttt{timestamptz} set by specific time intervals:
\begin{lstlisting}[language=SQL,numbers=none]
SELECT shiftScale(tstzset '{2025-01-01, 2025-01-02, 
  2025-01-03}', '1 day', '1 hour');
-- {"2025-01-02 00:00:00+00", "2025-01-02 00:30:00+00", "2025-01-02 01:00:00+00"}
\end{lstlisting}
    \item Transform a geometry set to a different spatial reference identifier:
\begin{lstlisting}[language=SQL,numbers=none]
SELECT asEWKT(transform( 
  geomset 'SRID=4326;{Point(2.340088 49.400250), 
  Point(6.575317 51.553167)}', 3812), 6);
-- SRID=3812;{"POINT(502773.429981 511805.120402)", "POINT(803028.908265 751590.742629)"}
\end{lstlisting}
    \item Expand the spatial dimension of a spatiotemporal bounding box by a value:
\begin{lstlisting}[language=SQL,numbers=none]
SELECT expandSpace(stbox 'STBOX XT(((1.0,2.0),
  (1.0,2.0)),[2025-01-01,2025-01-01])', 2.0);
-- STBOX XT(((-1,0),(3,4)),[2025-01-01 00:00:00+00, 2025-01-01 00:00:00+00])
\end{lstlisting}
    \item Expand the temporal dimension of a bounding box by an interval:
\begin{lstlisting}[language=SQL,numbers=none]
SELECT expandTime(tbox 'TBOXFLOAT XT([1.0,2.0],
  [2025-01-01,2025-01-02])', interval '1 day');
-- TBOXFLOAT XT([1, 2],[2024-12-31 00:00:00+00, 2025-01-03 00:00:00+00])
\end{lstlisting}
    \item Create a temporal geometry with a point, time span, and step interpolation:
\begin{lstlisting}[language=SQL,numbers=none]
SELECT asEWKT(tgeometry('Point(1 1)',
  tstzspan '[2025-01-01, 2025-01-02]', 'step'));
-- [POINT(1 1)@2025-01-01 00:00:00+00, POINT(1 1)@2025-01-02 00:00:00+00]
\end{lstlisting}
    \item Check if a temporal point geometry overlaps with a spatiotemporal bounding box:
\begin{lstlisting}[language=SQL,numbers=none]
SELECT tgeompoint '{[Point(1 1)@2025-01-01,
  Point(2 2)@2025-01-02, Point(1 1)@2025-01-03],
  [Point(3 3)@2025-01-04, Point(3 3)@2025-01-05]}'
  && stbox 'STBOX X((10.0,20.0),(10.0,20.0))';
-- false
\end{lstlisting}
    \item Restrict a temporal point geometry to a \texttt{timestamptz} span:
\begin{lstlisting}[language=SQL,numbers=none]
SELECT asText(atTime(tgeompoint 
  '{[Point(1 1)@2025-01-01, Point(2 2)@2025-01-02, 
  Point(1 1)@2025-01-03],[Point(3 3)@2025-01-04, 
  Point(3 3)@2025-01-05]}', 
  tstzspan '[2025-01-01,2025-01-02]'));
-- {[POINT(1 1)@2025-01-01 00:00:00+00, POINT(2 2)@2025-01-02 00:00:00+00]}
\end{lstlisting}
\end{itemize}
    
\section{Indexing System}
MobilityDuck implements an R-tree indexing system on the \texttt{stbox} (spatiotemporal bounding box) data type. R-trees are specifically designed for multidimensional data and provide efficient spatial access methods for indexing geographic and spatiotemporal information by organizing data using topological containment relations, making them ideal for spatial queries \cite{MobilityDataScience2025}. The indexing system integrates with DuckDB's query optimizer to enable efficient spatial query processing. Index scans are registered as specialized operators on \texttt{stbox} data. The indexing system seamlessly integrates with MEOS spatial functions, ensuring that:
\begin{itemize}
    \item Spatiotemporal bounding boxes are correctly extracted from temporal geometries,
    \item R-tree insertion and search operations use MEOS spatial predicates,
    \item Index maintenance operations preserve spatial integrity, and
    \item Query results are consistent with MobilityDB semantics.
\end{itemize}
\subsection{Index Registration}
To register the RTree index with DuckDB, the \texttt{RegisterRTreeIndex()} function configures the index type with its creation callbacks:
\begin{lstlisting}[language=C++,numbers=none]
void RTreeModule::RegisterRTreeIndex(
    DatabaseInstance &db) {
  IndexType index_type;
  index_type.name = RTreeIndex::TYPE_NAME;
  index_type.create_instance = RTreeIndex::Create;
  index_type.create_plan = RTreeIndex::CreatePlan;
  db.config.GetIndexTypes().RegisterIndexType(index_type);
}
\end{lstlisting}
The \texttt{Create} and \texttt{CreatePlan} methods are essential callbacks that DuckDB uses to instantiate the index and generate execution plans. The \texttt{TYPE\_NAME} is defined as \texttt{TRTREE} to avoid naming conflicts with the \texttt{RTREE} index type already present in DuckDB's Spatial extension.

\subsection{Index Construction }
Our implementation supports two distinct scenarios for index construction, each optimized for different use cases in database operations.

\subsubsection{Incremental Construction: Index-First Approach}
    
In the first scenario, an index already exists on a table, and new data is being inserted. When new data is inserted into a table that already has an RTree index, the \texttt{Append} method handles incremental updates:
\begin{lstlisting}[language = C++,numbers=none]
ErrorData RTreeIndex::Append(IndexLock &lock, DataChunk 
    &appended_data, Vector &row_identifiers) {
  DataChunk expression_result;
  expression_result.Initialize(
    Allocator::DefaultAllocator(), logical_types);
  ExecuteExpressions(appended_data, expression_result);
  Construct(expression_result, row_identifiers);
  return ErrorData();
}    
\end{lstlisting}
This method evaluates index expressions on the new data and constructs index entries using the MEOS RTree insertion functionality. The \texttt{Construct} method processes data chunks and inserts them into the RTree structure. Once the \texttt{stbox} is prepared, the method applies the MEOS library's insert function \texttt{rtree\_insert} to handle the actual insertion into the RTree data structure. 

\subsubsection{Bulk Construction: Data-First Approach}
The second scenario occurs when creating an index on a table that already contains data, typically through a \texttt{CREATE INDEX} statement. This situation requires a different strategy optimized for processing large volumes of existing data.
Our implementation follows a three-phase pipeline that leverages parallel processing for efficiency:

\begin{itemize}
    \item \textbf{Phase 1: Data Collection\hspace{1mm}} As DuckDB's execution framework scans the table in parallel across multiple threads, each thread processes its assigned data partition through the \texttt{Sink()} method. This method receives chunks of data containing \texttt{stbox} values and row identifiers, appending them to thread-local storage.
    \item \textbf{Phase 2: Data Combination\hspace{1mm}} The \texttt{Combine()} method consolidates thread-local collections into a single global dataset through thread-safe merging operations. This consolidation is protected by a mutex to ensure data consistency.
    \item \textbf{Phase 3: Index Construction\hspace{1mm}} In this phase, the system constructs the actual index entries from the collected data. For each chunk, the task deserializes \texttt{stbox} data, performs SRID normalization, and collects valid entries into arrays. It then calls the \texttt{BulkConstruct} method with these arrays, which inserts entries through \texttt{rtree\_insert}.
\end{itemize}

\subsection{Query Optimization and Index Scan Injection}
DuckDB's query optimizer automatically replaces sequential scans with index scans when applicable predicates are detected. To enable this optimization, the index registers a scan matcher for operators between two \texttt{stbox} operands. When a spatial filter predicate matches the indexable pattern, the optimizer substitutes the original table scan with an index scan operator.
MobilityDuck currently supports pattern matching for the spatial overlap operator (\texttt{\&\&}) between two \texttt{stbox} operands. During query optimization, when the optimizer encounters a filter expression containing this operator, the index attempts to bind the operands. If one operand is a constant \texttt{stbox} value, the index can perform an efficient bounding-box search using the R-tree structure.

During index scan execution, the scan normalizes the query's spatial reference system (SRID) to ensure geometric consistency. Then, the normalized bounding box queries the R-tree structure using the underlying MEOS R-tree implementation, which returns identifiers of all entries whose bounding boxes overlap with the query region. Finally, the scan operator iterates through these candidate row IDs, retrieving and returning qualifying tuples to the query pipeline.

\subsection{Indexing Example}
This section demonstrates the use of the implemented R-tree index on the \texttt{stbox} type. This example follows the incremental construction scenario previously discussed, where the index is created first on a table and new tuples are inserted afterwards.

First, a test table \texttt{test\_geo} is created with two simple columns, \texttt{times} (of type \texttt{timestamptz}) and \texttt{box} (of type \texttt{stbox}):

\begin{lstlisting}[language=SQL,numbers=none]
CREATE TABLE test_geo(
  "times" timestamptz,
  "box" stbox
);
\end{lstlisting}

Next, an R-tree index is created on the \texttt{box} column:

\begin{lstlisting}[language=SQL,numbers=none]
CREATE INDEX rtree_stbox ON test_geo USING TRTREE(box);
\end{lstlisting}

Then, new tuples are inserted into the table. Here, we use a script to insert synthetic data:
\begin{lstlisting}[language=SQL,numbers=none]
INSERT INTO test_geo 
SELECT ('2025-08-11 12:00:00'::timestamp + 
    INTERVAL (i || ' minutes')) AS times,
  ('STBOX X((' || 
    (i * 1.0)::DECIMAL(10,2) || ',' ||        
    (i * 1.0)::DECIMAL(10,2) || '),(' || 
    (i * 1.0 + 0.5)::DECIMAL(10,2) || ',' || 
    (i * 1.0 + 0.5)::DECIMAL(10,2) || '))') AS stbox_data
FROM generate_series(1, 1000) AS t(i);
\end{lstlisting}

\begin{figure}[h!]
    \centering
    \includegraphics[width=0.4\linewidth]{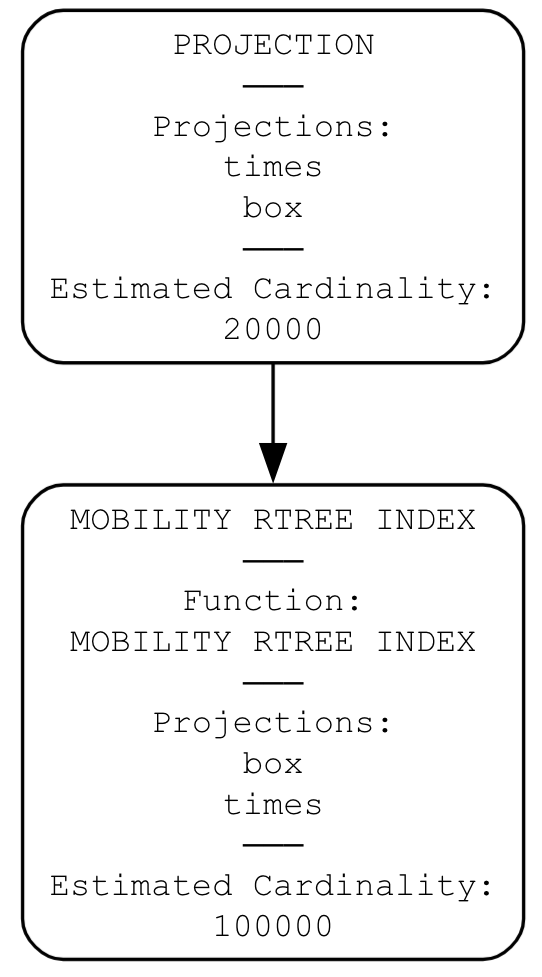}
    \caption{Execution plan of query}
    \label{fig:index}
\end{figure}

To test the newly created R-tree index, we use the following query which includes a \texttt{WHERE} clause and utilizes the overlaps (\texttt{\&\&}) predicate:
\begin{lstlisting}[language=SQL,numbers=none]
SELECT * FROM test_geo 
WHERE box && STBOX('STBOX X((1000.0,1000.0),
  (1100.0,1100.0))');
\end{lstlisting}

The execution plan that DuckDB generates for this query is shown in Figure~\ref{fig:index}.

\begin{figure}[h!]
    \centering
    \includegraphics[width=0.47\textwidth]{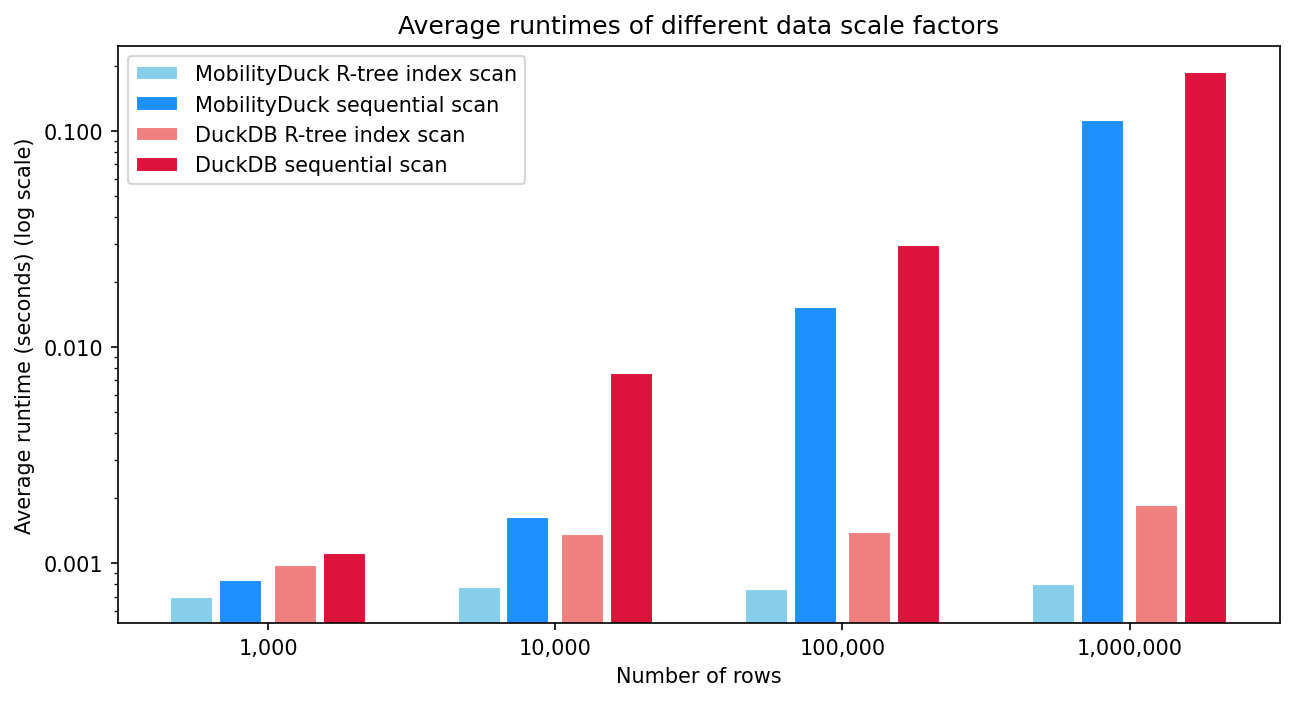}
    \caption{R-tree index scan versus sequential scan of MobilityDuck and DuckDB on different scale factors}
    \label{fig:index_scales}
\end{figure}

Using this query, we can compare the performances of MobilityDuck R-tree index scan and sequential scan. Furthermore, we compare MobilityDuck with native DuckDB R-tree index scan, supported by the Spatial extension.\footnote{\url{https://duckdb.org/docs/stable/core_extensions/spatial/r-tree_indexes.html}} We create a second table for this test, called \texttt{test\_geo\_geom}, which is similar to \texttt{test\_geo} with the addition of the column \texttt{geom} of time \texttt{geometry}. We insert synthetic data into the \texttt{times} and \texttt{box} columns like above, before updating the table to insert geometry values into the column \texttt{geom}, then create an R-tree index on this column:
\begin{lstlisting}[language=SQL,numbers=none]
UPDATE test_geo_geom SET geom = geometry(box)::GEOMETRY;
CREATE INDEX rtree_geom ON test_geo_geom
  USING RTREE (geom);
\end{lstlisting}

We test this index using a modified version of the overlap query:
\begin{lstlisting}[language=SQL,numbers=none]
SELECT * FROM test_geo_geom 
WHERE ST_Intersects(geom, {min_x: 1000, min_y:1000,
  max_x: 1100, max_y: 1100}::BOX_2D);
\end{lstlisting}

Figure~\ref{fig:index_scales} showcases the performances of MobilityDuck and native DuckDB with R-tree indexes (index on \texttt{stbox} for MobilityDuck and on \texttt{geometry} for DuckDB) and with sequential scans. Comparisons are shown for 4 different scale factors, corresponding to 4 different sizes of the \texttt{test\_geo} and \texttt{test\_geo\_geom} tables tuned by changing the second argument of the \texttt{generate\_series()} functions: 1,000 rows, 10,000 rows, 100,000 rows, and 1,000,000 rows. The y-axis represents the average runtime, in seconds, taken over 5 test runs and is shown in logarithmic scale. For both MobilityDuck and plain DuckDB, the runtime of sequential scan grows fast proportional to the sizes of the tables. Meanwhile, both implementations of R-tree index show little increase in runtime as the tables grow in size. Compared to DuckDB's, MobilityDuck's R-tree index scan shows better runtime especially in the larger scale factors, performing virtually the same across all 4 scales (0.0007 seconds for 1,000 rows, 0.00078 seconds for 10,000 rows, 0.00076 seconds for 100,000 rows, and 0.0008 seconds for 1,000,000 rows).

\section{BerlinMOD-Hanoi }
\label{sec:berlinmod-hanoi}
BerlinMOD is the de facto benchmark for evaluating spatiotemporal database systems. However, its MobilityDB implementation tailors the road network and mobility model to Brussels, Belgium. To extend its applicability, we created \textbf{BerlinMOD-Hanoi}, an adaptation of the benchmark to the urban setting of Hanoi, Vietnam. This allows us to evaluate MobilityDuck on realistic mobility data from a non-European city with different traffic patterns, densities, and cultural mobility habits. BerlinMOD-Hanoi datasets, SQL scripts, and visualization functions are publicly available.\footnote{\url{https://github.com/MobilityDB/MobilityDB-BerlinMOD-Hanoi}}

\subsection{Dataset Preparation}
We followed the BerlinMOD methodology but replaced Brussels’s road network with the one from OpenStreetMap (OSM) for Hanoi:
\begin{itemize}
    \item Extracted the Hanoi road network using \texttt{osm2pgsql} and \texttt{osm2pgrouting}, configured with BerlinMOD’s \texttt{mapconfig.xml} to select road types.  
    \item Constructed a routable network topology with pgRouting.  
    \item Applied the BerlinMOD trip generation logic, adjusted with population statistics of Hanoi’s administrative regions using a customized SQL script \texttt{hanoi\_preparedata.sql}.  
\end{itemize}

The generated trips simulate commuting activities, sampled according to home–work distributions derived from administrative region statistics. Each trip is represented as a temporal sequence of positions (\texttt{tgeompoint}) with associated time instants, fully compatible with MEOS/MobilityDB types.

\subsection{Dataset Characteristics}
BerlinMOD-Hanoi produces scalable datasets through the scale factor (SF) parameter. Table~\ref{tab:hanoi-datasets} summarizes the datasets we generated and released.


\begin{table}[h]
\centering
\caption{BerlinMOD-Hanoi datasets at different scale factors (SF). Each dataset contains synthetic trips generated on the Hanoi road network using OSM data.}
\label{tab:hanoi-datasets}
\begin{tabular}{ccccc}
\toprule
Scale Factor & Vehicles & Days & Trips & Size \\
\midrule
SF 0.01 & 200  & 5  & 2,903  & 214.6 MB \\
SF 0.02 & 283  & 6  & 4,641  & 310.4 MB \\
SF 0.05 & 447  & 8  & 9,491  & 626.6 MB \\
SF 0.1  & 632  & 11 & 18,910 & 1.27 GB \\
\bottomrule
\end{tabular}
\end{table}

\begin{figure*}[t]
\centering
\begin{minipage}[t]{0.45\textwidth}
  \centering
  \includegraphics[width=\linewidth]{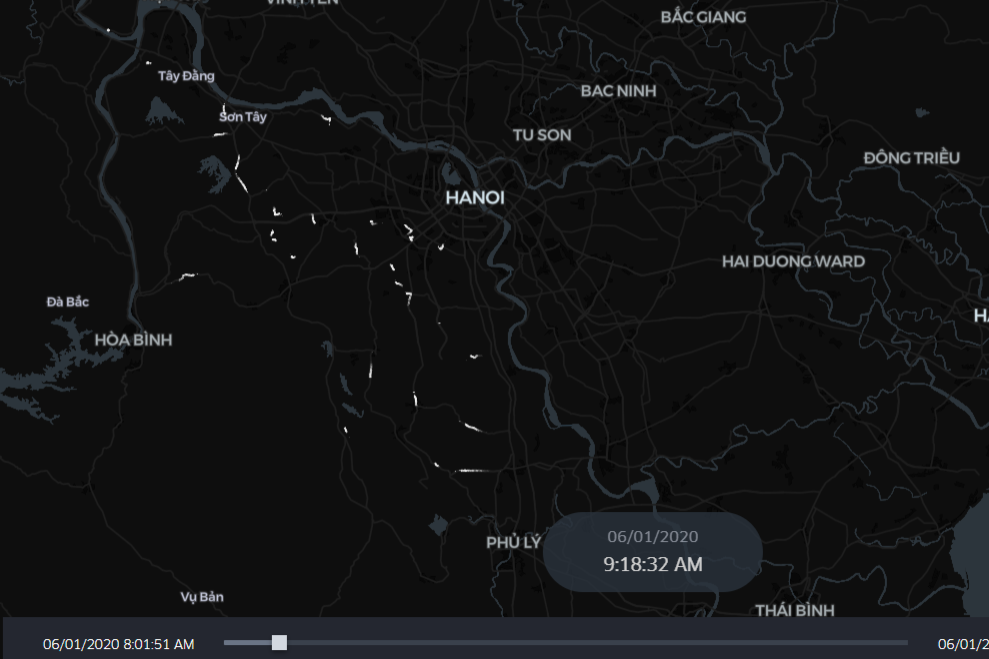}
  \captionof{figure}{Visualization of animated synthetic trips in Hanoi generated by BerlinMOD-Hanoi.}
  \label{fig:kepler-hanoi-trips}
\end{minipage}\hfill
\begin{minipage}[t]{0.45\textwidth}
  \centering
  \includegraphics[width=\linewidth]{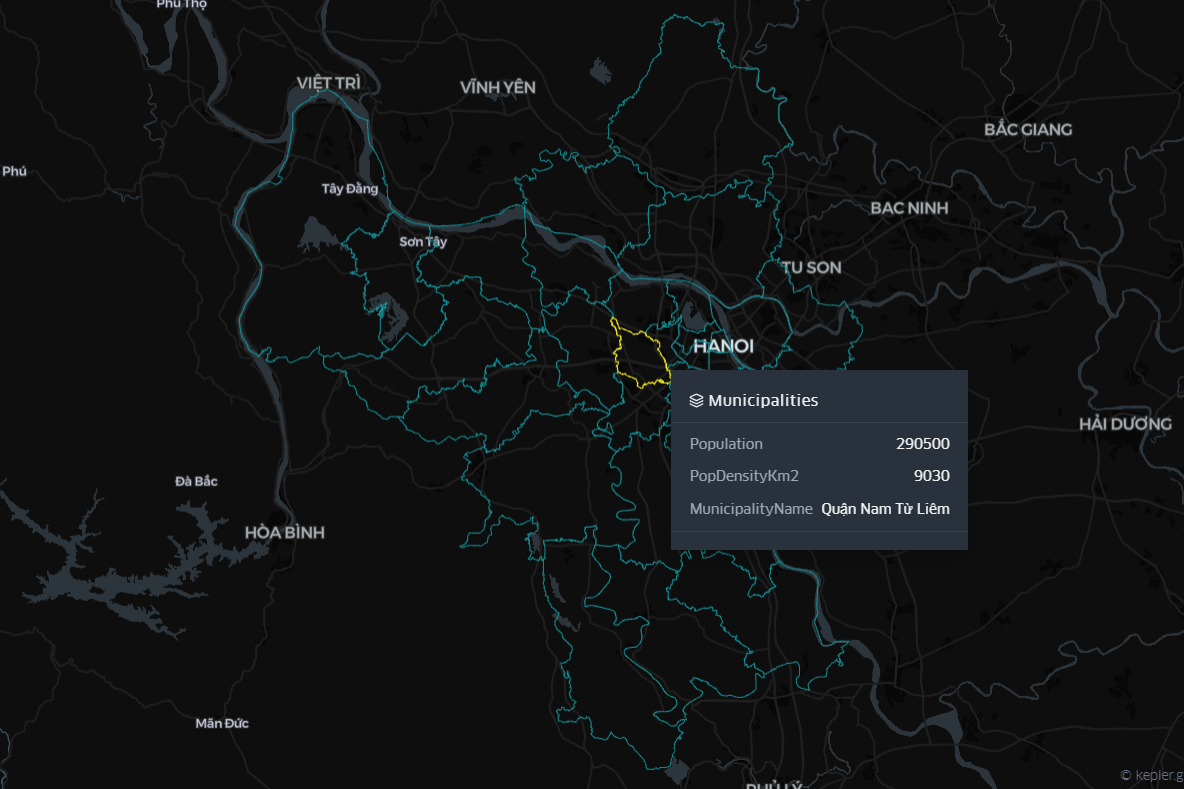}
  \captionof{figure}{Hanoi administrative regions used in BerlinMOD-Hanoi.}
  \label{fig:hanoi-municipalities}
\end{minipage}
\end{figure*}

We also provide \textbf{GeoJSON exports} of trips and administrative regions, enabling visualization in \textbf{Kepler.gl}.\footnote{\href{https://kepler.gl/}{https://kepler.gl/}} Figure~\ref{fig:kepler-hanoi-trips} shows an animation of the synthetic trips generated by BerlinMOD-Hanoi, while Figure~\ref{fig:hanoi-municipalities} shows the administrative boundaries used to sample realistic home and work locations. 

\begin{figure}[b]
\centering
\includegraphics[width=0.95\linewidth]{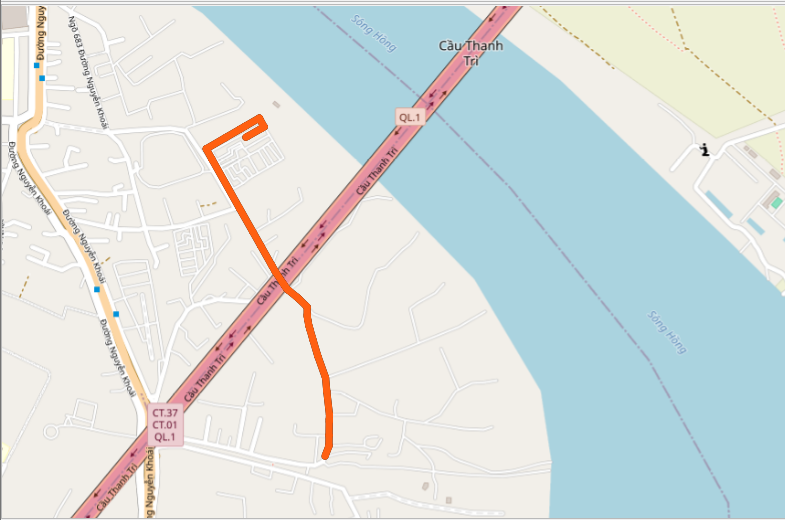}
\caption{Visualization of a synthetic trip in Hanoi generated by BerlinMOD-Hanoi (QGIS+MOVE plugin).}
\label{fig:QGIS+move}
\end{figure}

Another possible way to visualize the data is to use traditional tool QGIS\footnote{\href{https://qgis.org/}{https://qgis.org/}} and the MOVE\footnote{\href{https://github.com/MobilityDB/move}{https://github.com/MobilityDB/move}} plugin. Figure~\ref{fig:QGIS+move} shows result of query databases using \texttt{SELECT} statements as QGIS layers.


\section{Experimental Evaluation}
This section demonstrates the utilization of MobilityDuck in trajectory manipulation using the BerlinMOD-Hanoi dataset. Initial data exploration is conducted by integrating MobilityDuck with DuckDB's Python API and traditional Python libraries for visualization. Additionally, MobilityDuck (DuckDB) is evaluated against MobilityDB (PostgreSQL) using 17 range queries provided by the BerlinMOD benchmark.

\subsection{Experimental Setup}
All experiments in this section were conducted on an Oracle virtual machine running Ubuntu 20.04, 3 CPUs, 18GB RAM. The BerlinMOD-Hanoi datasets used were generated at 4 different scale factors (see Section~\ref{benchmark-intro}). MobilityDuck was built with DuckDB version 1.3.2. The exploratory steps using DuckDB's Python API were run with Python 3.9, DuckDB Python client 1.3.2. MobilityDB benchmarking was conducted on PostgreSQL 15.13.

\subsection{Use Case Demonstration}
This section presents a preliminary demonstration to showcase MobilityDuck's capacity in manipulating spatiotemporal data as well as its potential in integrating with other tools, which in this case include DuckDB's Python API and other Python libraries traditionally used for analyzing and visualizing data.

The demonstration utilizes the existing BerlinMOD-Hanoi dataset containing trips, where each row reveals the coordinates (longitude and latitude) of a specific vehicle made during a given trip at a specific timestamp. The coordinates and timestamp of a row are used to create a \texttt{tgeompoint} value, which is ideal for representing a temporal geometry (which, in this case, is a \texttt{POINT}). Then, the \texttt{tgeompoint} values are aggregated by vehicle IDs and trip IDs to create \texttt{tgeompointSeq} values, which are still temporal geometries with the additional sequence subtype to represent the evolution of the geometries over a sequence of time instants. Finally, to facilitate the visualization process in Python, the \texttt{tgeompointSeq} values are turned into trajectories in \texttt{GEOMETRY} type using the \texttt{trajectory()} function.

The \texttt{GEOMETRY} data type, which is prevalent in other spatial database systems such as PostgreSQL (extended with PostGIS), is supported in DuckDB by the Spatial extension.\footnote{\href{https://duckdb.org/docs/stable/core_extensions/spatial/overview.html}{https://duckdb.org/docs/stable/core\_extensions/spatial/overview.html}} As such, handling the \texttt{GEOMETRY} type is beyond the scope of MobilityDuck. The latest iteration of MobilityDuck includes a preliminary interface with Spatial's \texttt{GEOMETRY} and \texttt{WKB\_BLOB} types to ensure the usability of MEOS functions originally involving geometries. When integrating with Python, the geometries can be loaded using the Shapely library\footnote{\href{https://shapely.readthedocs.io/en/stable/}{https://shapely.readthedocs.io/en/stable/}} to return a GeoPandas\footnote{\href{https://geopandas.org/en/stable/}{https://geopandas.org/en/stable/}} dataframe for further processing and visualizing.

\begin{figure*}[h!]
\centering
    \begin{minipage}[b]{0.225\textwidth}
        \includegraphics[width=\textwidth]{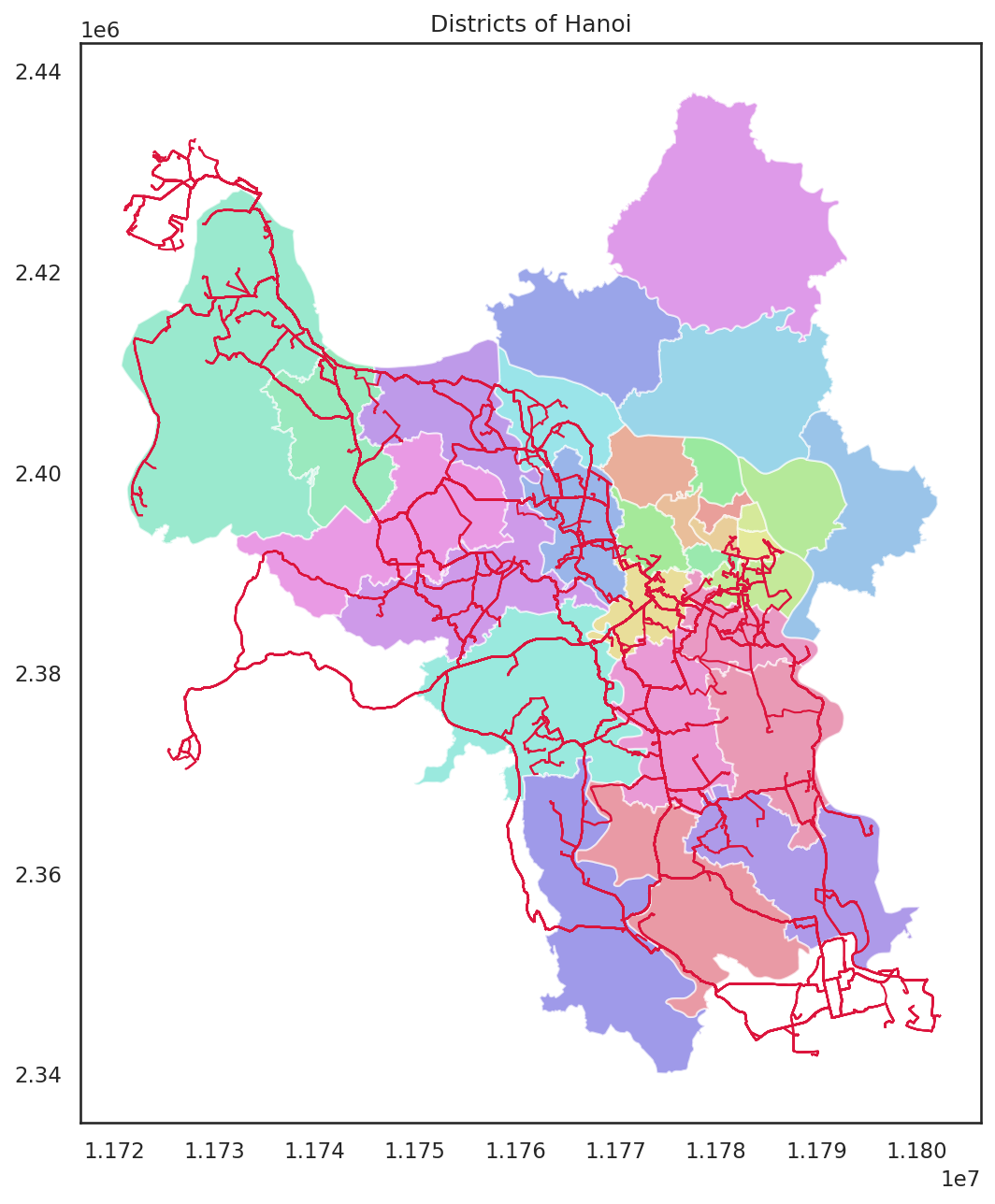}
        \caption{Trajectories of all trips\\}
        \label{fig:fig0}
    \end{minipage}
    \hspace{1mm}
    \begin{minipage}[b]{0.225\textwidth}
        \includegraphics[width=\textwidth]{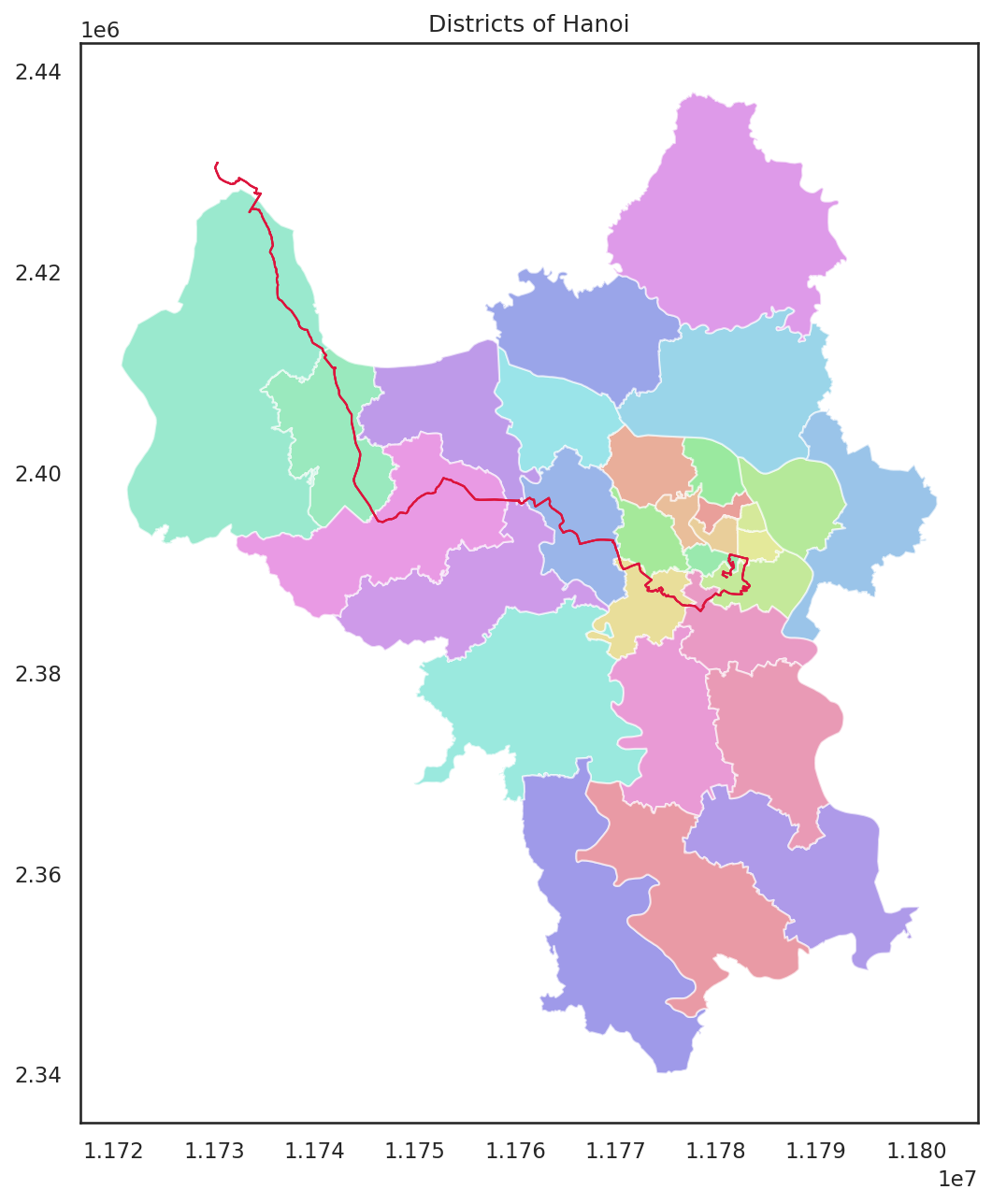}
        \caption{Trajectory of the trip crossing the highest number of districts}
        \label{fig:fig1}
    \end{minipage}
    \hspace{1mm} 
    \begin{minipage}[b]{0.225\textwidth}
        \includegraphics[width=\textwidth]{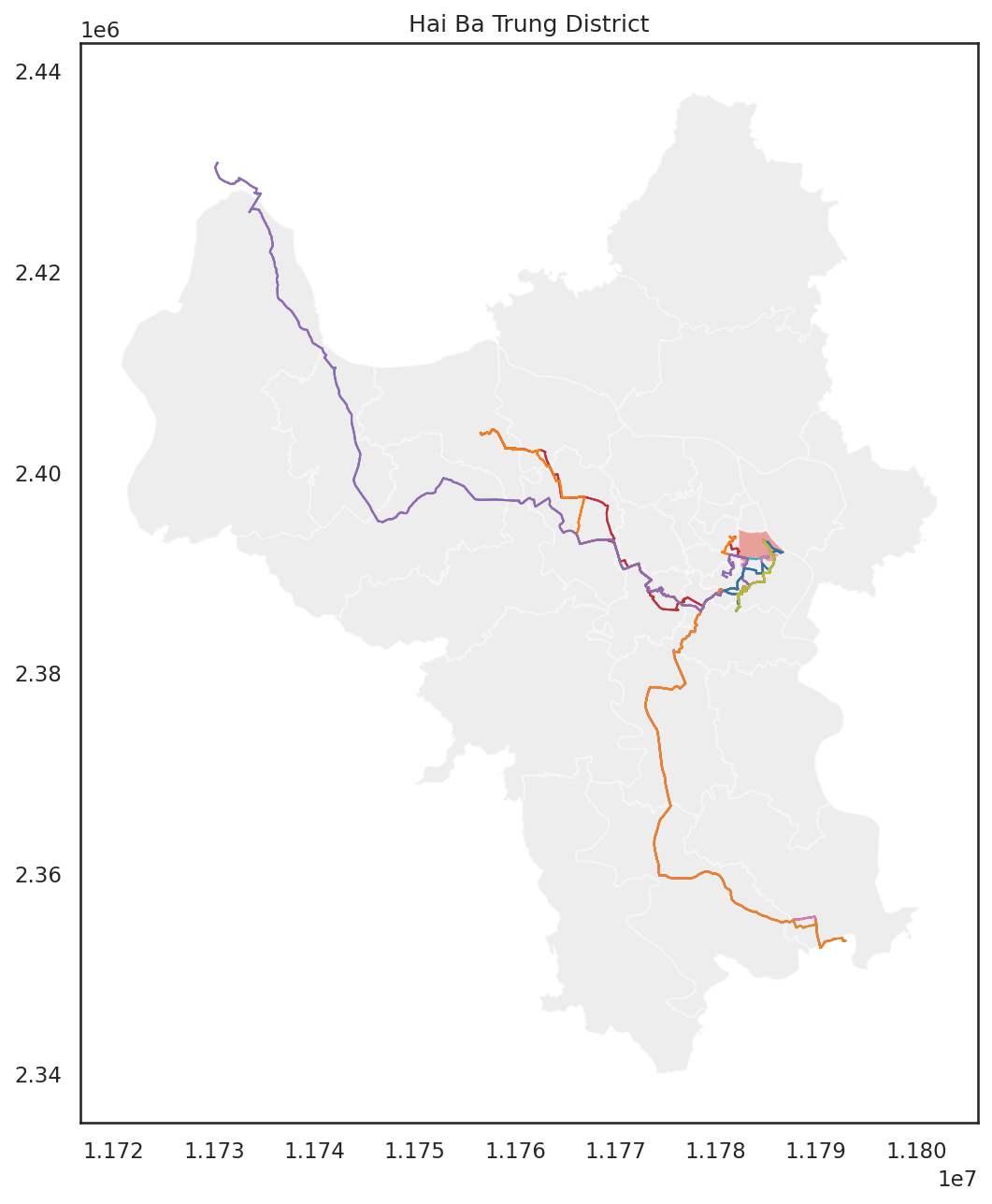}
        \caption{Trips crossing the Hai Ba Trung District \\}
        \label{fig:fig2}
    \end{minipage}
    \hspace{1mm}
    \begin{minipage}[b]{0.27\textwidth}
        \includegraphics[width=\textwidth]{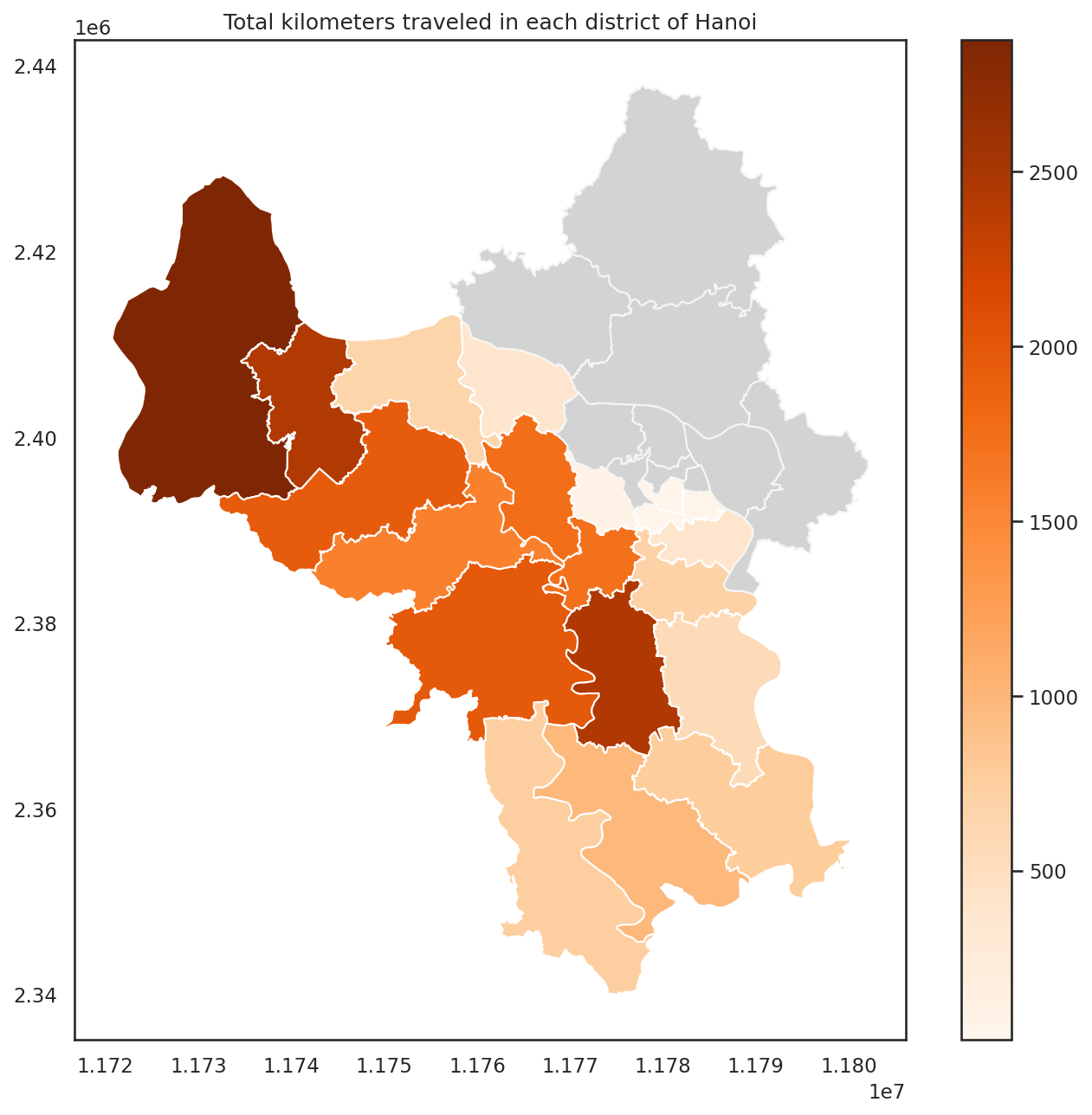}
        \caption{Districts by total distance traveled \\}
        \label{fig:fig3}
    \end{minipage}
\end{figure*}

\begin{figure}[h!]
    \centering
    \includegraphics[width=0.4\textwidth]{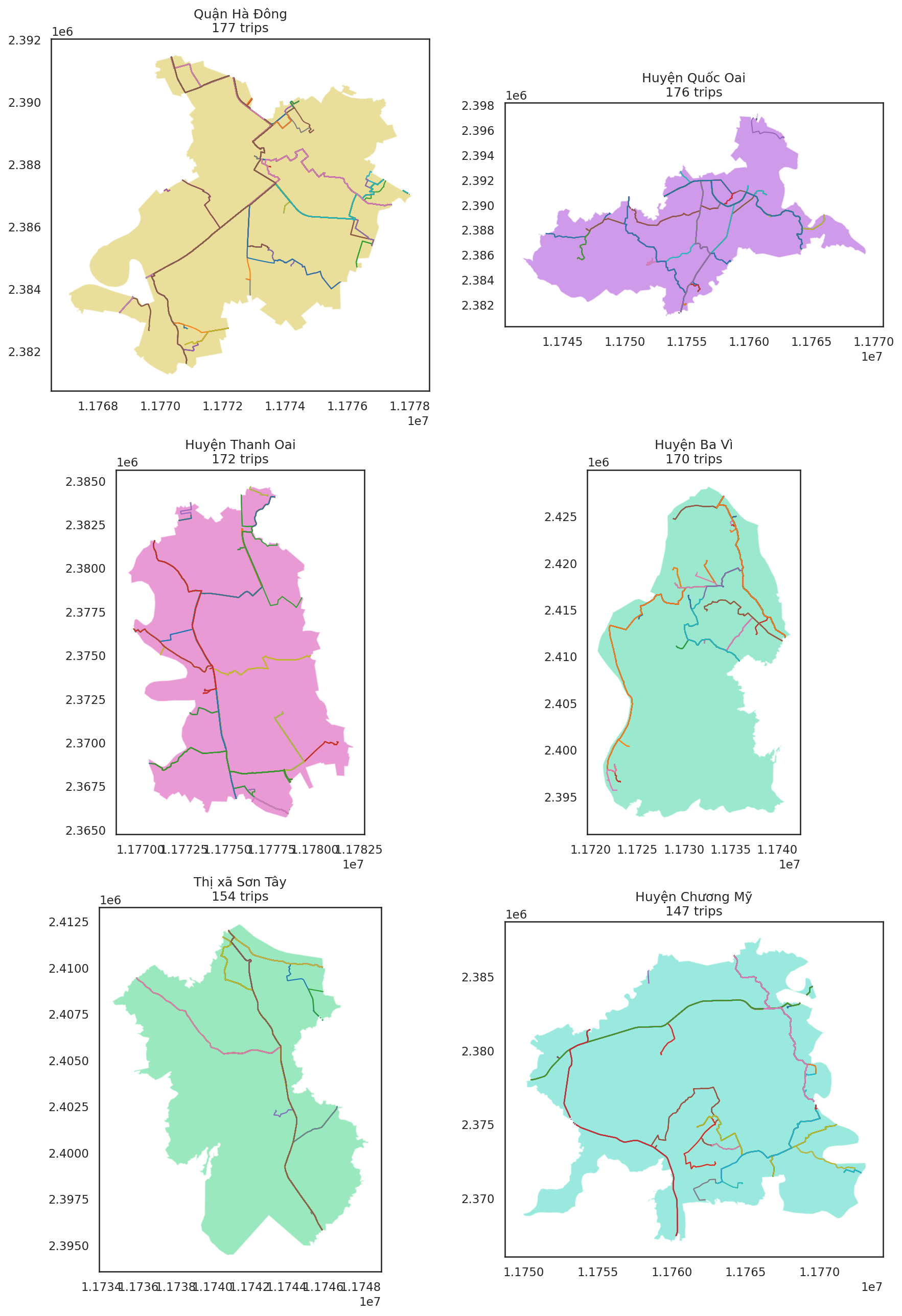}
    \caption{Top 6 districts with highest numbers of trips crossing; trips are clipped to districts}
    \label{fig:fig4}
\end{figure}

\begin{figure}[h!]
    \centering
    \begin{subfigure}{0.21\textwidth}
        \centering
        \includegraphics[width=\textwidth]{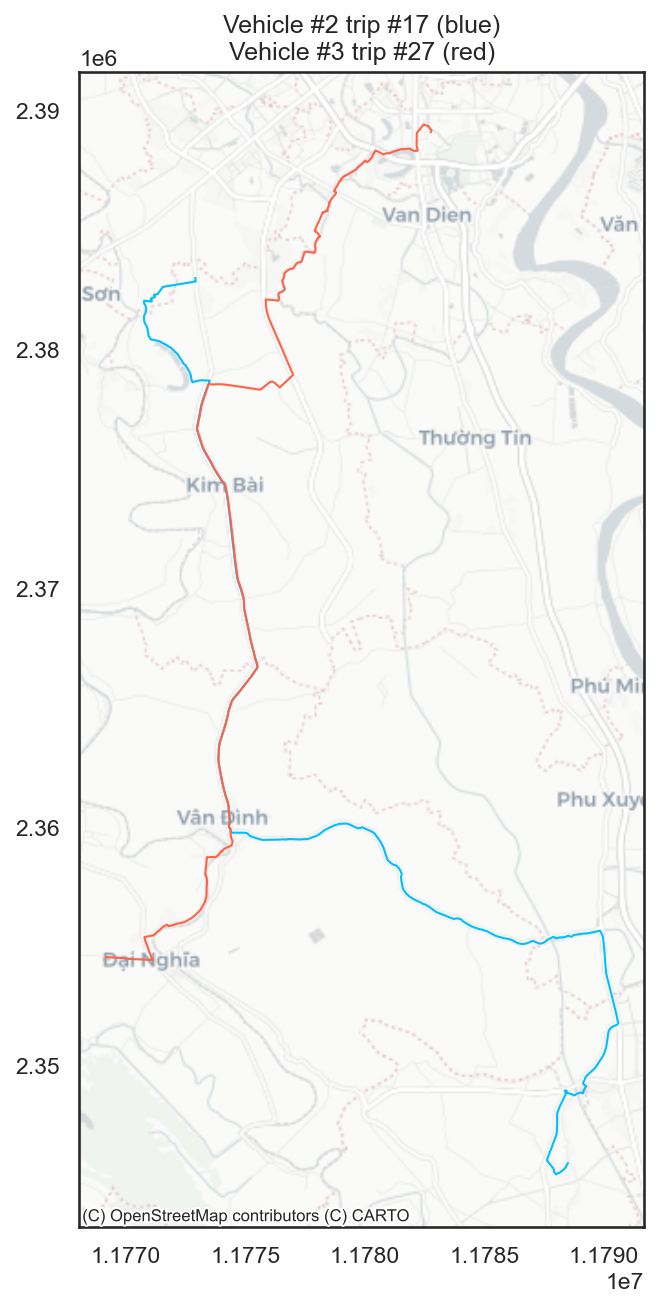}
    \end{subfigure}
    \begin{subfigure}{0.236\textwidth}
        \centering
        \includegraphics[width=\textwidth]{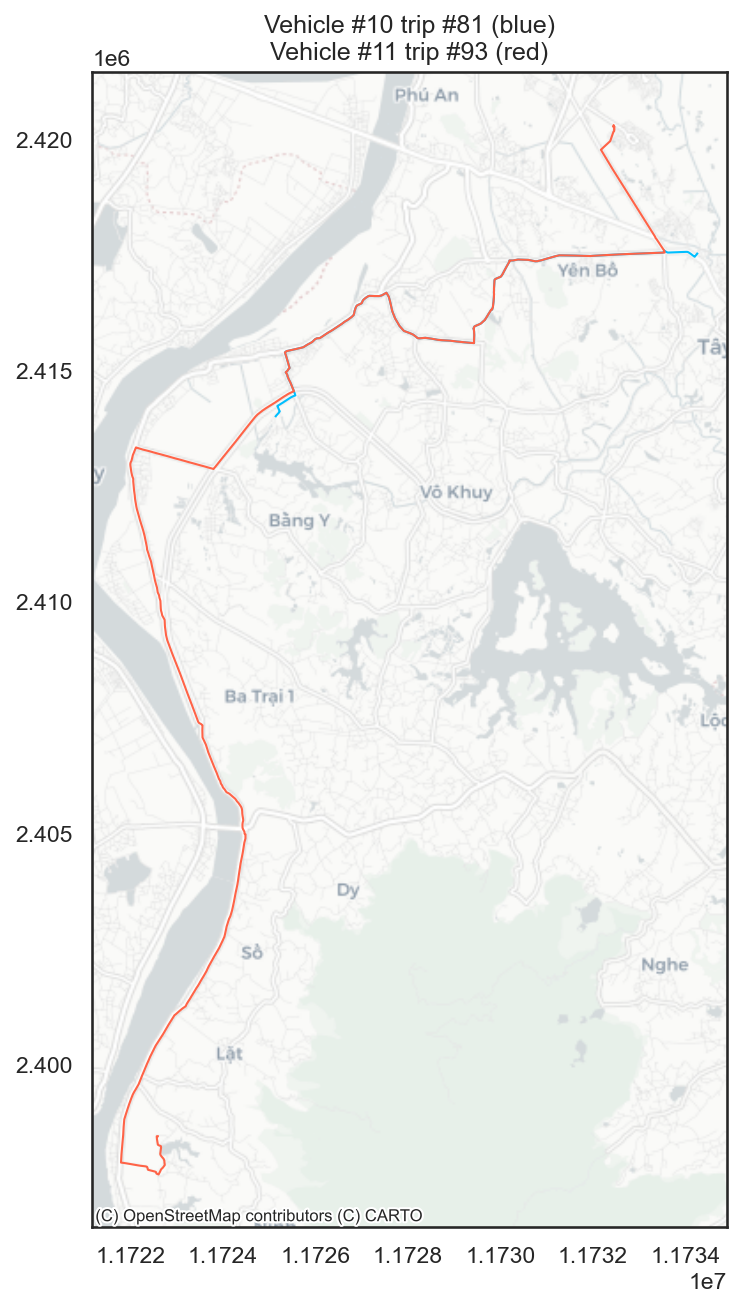}
    \end{subfigure}
    \caption{Trips made by 2 pairs of vehicles that have ever been as close to each other as 10 meters or under}
    \label{fig:fig5}
\end{figure}

Having loaded the data and conducted the aforementioned data preparation steps, a number of operations are run and their results are captured and visualized:
\begin{enumerate}
    \item Show the trajectories of all trips (Figure~\ref{fig:fig0})
    \item Show the trip(s) that cross the highest number of districts (Figure~\ref{fig:fig1})
    \item Show the trips that cross Hai Ba Trung district (Figure~\ref{fig:fig2})
    \item Show the total distance traveled per district (Figure~\ref{fig:fig3})
    \item Show 6 districts with the highest number of trips crossing them, and show parts of the trips that cross the districts (Figure~\ref{fig:fig4})
    \item Show trips made by pairs of vehicles that have ever been as close to each other as 10 meters or under (Figure~\ref{fig:fig5})
\end{enumerate}



The SQL queries and Python script for recording and visualizing results are available as a Jupyter Notebook in the example section of MobilityDuck repository.

Below, we show the SQL query used for operation (4):
\begin{lstlisting}[language=SQL,numbers=none]
SELECT h.municipalityname, round(
  ( sum(length(atGeometry(t.trip, h.geom::WKB_BLOB)) ) /
    1000)::numeric, 3) AS total_km
FROM trajectories t, hanoi h
WHERE ST_Intersects(t.traj, h.geom)
GROUP BY h.municipalityname;
\end{lstlisting}

The \texttt{trajectories} table contains the processed trip data, where the \texttt{trip} column is of type \texttt{tgeompoint}. The \texttt{hanoi} table contains data on the district-level subdivisions of the city of Hanoi. The \texttt{geom} column of this table contains the geometry of the districts and towns comprising the city (in Figures~\ref{fig:fig0} and~\ref{fig:fig1}). The \texttt{atGeometry(tgeom, geometry)} restricts the temporal geometry (first argument) to the geometry (second argument), returning a \texttt{tgeompoint} value representing the parts of the given trip that were traveled within a certain district. Here, the \texttt{geom} value from the table \texttt{hanoi} is explicitly cast as a \texttt{WKB\_BLOB} before passing to \texttt{atGeometry}. This is due to the actual implementation of the function that expects a \texttt{WKB\_BLOB} to later cast into a geometry value. The returned value is then passed to the \texttt{length(tgeom)} function which returns the length traversed by the temporal point. The condition in the \texttt{WHERE} clause quickly filters trips crossing a given district using the \texttt{ST\_Intersects()} function from DuckDB's Spatial. The first argument of this function comes from \texttt{trajectories}'s \texttt{traj} column, a \texttt{geometry} column representing the full trajectories of trips.

The SQL query used for operation (6) is shown next:
\begin{lstlisting}[language=SQL,numbers=none]
SELECT DISTINCT t1.VehicleId AS VehicleId1,
  t1.TripId AS TripId1, ST_AsText(t1.Traj) AS Traj1,
  t2.VehicleId AS VehicleId2, t2.TripId AS TripId2,
  ST_AsText(t2.Traj) AS Traj2,
FROM (SELECT * FROM trajectories t1 LIMIT 100) t1,
  (SELECT * FROM trajectories t2 LIMIT 100) t2
WHERE t1.VehicleId < t2.VehicleId AND
  eDwithin(t1.Trip, t2.Trip, 10.0)
ORDER BY t1.VehicleId, t2.VehicleId;
\end{lstlisting}

The \texttt{eDwithin()} function used in the second join condition belongs to the group of relationships that can be used to determine whether a specific topological or distance relationship is ever/always satisfied, such as \texttt{eIntersects()} (if two entities ever intersect) and \texttt{aTouches()} (if two entities always touch). In this case, \texttt{eDwithin()} is called to determine if the two trips have ever been within 10 meters of one another.

\subsection{BerlinMOD-Hanoi Benchmarking}\label{benchmark-results}
\subsubsection{Introduction}\label{benchmark-intro}
Performances of MobilityDuck (in DuckDB) and MobilityDB (in PostgreSQL) are compared using 17 range-style queries on the BerlinMOD-Hanoi dataset of 4 scale factors: SF-0.001, SF-0.002, SF-0.005, and SF-0.01. Table~\ref{tab:benchmark-sf} summarizes the scale factors. Under constraints of computational resources, the benchmarking is conducted for these small scale factors. The next iterations of the benchmark in the near future will assess the system using larger-scale datasets, such as the scale factors presented in Table~\ref{tab:hanoi-datasets}, to gain better insights into the performance and scalability of MobilityDuck.

\begin{table}[h]
\centering
\caption{BerlinMOD-Hanoi datasets at 4 different scale factors (SF) used for the benchmark}
\label{tab:benchmark-sf}
\begin{tabular}{ccc}
\toprule
Scale factor & Number of vehicles & Number of trips \\ \midrule
SF-0.001              & 63                   & 549               \\
SF-0.002              & 89                   & 758               \\
SF-0.005              & 141                  & 1620              \\
SF-0.01               & 200                  & 2903              \\
\bottomrule
\end{tabular}
\end{table}

For MobilityDB on PostgreSQL, the queries are run twice, once without indexes, and once with a number of indexes on spatial, temporal, or spatiotemporal attributes.

On each tool, a number of tables are created and populated in advance. The loading phase is excluded from the evaluation, and only the elapsed times of running the queries are used for the subsequent comparisons. The business questions and corresponding SQL queries are available in the benchmark section of MobilityDuck repository. We introduce next a selected number of queries.

\vspace{1mm}\noindent
\textbf{Query 3:} \textit{Where have the vehicles with licenses from \texttt{Licenses1} been at each of the instants from \texttt{Instants1}?}

\begin{lstlisting}[language=SQL,numbers=none]
SELECT DISTINCT l.License, i.InstantId,
  i.Instant AS Instant,
  valueAtTimestamp(t.Trip, i.Instant)::GEOMETRY AS Pos
FROM Trips t, Licenses1 l, Instants1 i
WHERE t.VehicleId = l.VehicleId AND
  t.Trip::tstzspan @> i.Instant
ORDER BY l.License, i.InstantId;
\end{lstlisting}

The \texttt{Licenses1} table is a sample of 10 tuples extracted from the \texttt{Licenses} table, meant to keep the runtime of the query reasonable for demonstration. Each tuple in the table contains the license ID, license number, and the associated vehicle ID. The same applies for the \texttt{Instants1} table, which is an extracted sample from the \texttt{Instants} table containing timestamps with timezone. The \texttt{valueAtTimestamp()} function used in the \texttt{SELECT} statement takes in a temporal type as the first argument (such as a temporal integer \texttt{tint}, a temporal float \texttt{tfloat}, etc.) and a timestamp as the second argument and returns the value of the temporal argument (as the base type) at the given timestamp. In this query, the first argument of the function comes from the column \texttt{Trip} of type \texttt{tgeompoint} from the table \texttt{Trips}, yielding non-temporal geometry values (points, specifically). The second join condition in the \texttt{WHERE} utilizes the \texttt{@>} (contains) operator. The \texttt{Trip} value is first cast into the \texttt{tstzspan} type. This is one of the \texttt{span} template types used for representing ranges of values (see Table~\ref{typetable}). Specifically, \texttt{tstzspan} represents ranges of \texttt{timestamptz} values. Since \texttt{Trip} represents temporal geometric points, casting this into a \texttt{tstzspan} value essentially extracts the temporal span of a given trip, and the \textit{contains} (\texttt{@>}) predicate returns a boolean value representing whether this span contains a given instant (timestamp).

\vspace{1mm}\noindent
\textbf{Query 5:} \textit{What is the minimum distance between places, where a vehicle with a license from \texttt{Licenses1} and a vehicle with a license from \texttt{Licenses2} have been?}

\begin{lstlisting}[language=SQL,numbers=none]
WITH Temp1(License1, Trajs) AS (
  SELECT l1.License, ST_Collect(list(trajectory(t1.Trip)::GEOMETRY))
  FROM Trips t1, Licenses1 l1
  WHERE t1.VehicleId = l1.VehicleId
  GROUP BY l1.License ),
Temp2(License2, Trajs) AS (
  SELECT l2.License, ST_Collect(list(trajectory(t2.Trip)::GEOMETRY))
  FROM Trips t2, Licenses2 l2
  WHERE t2.VehicleId = l2.VehicleId
  GROUP BY l2.License )
SELECT License1, License2, ST_Distance(t1.Trajs, t2.Trajs) AS MinDist
FROM Temp1 t1, Temp2 t2  
ORDER BY License1, License2;
\end{lstlisting}

This query processes high volumes of trajectory values, which involves casting between \texttt{WKB\_BLOB} and \texttt{GEOMETRY} type, heavily increasing the runtimes. To optimize such bulky operations, we implemented MobilityDuck-native equivalents of DuckDB's spatial functions that take \texttt{GEOMETRY} as input, such as \texttt{ST\_Collect()} and \texttt{ST\_Distance()}. The modified version of the query is shown below:
\begin{lstlisting}[language=SQL,numbers=none]
WITH Temp1(License1, Trajs) AS (
  SELECT l1.License, 
    collect_gs(list(trajectory_gs(t1.Trip)))
  FROM Trips t1, Licenses1 l1
  WHERE t1.VehicleId = l1.VehicleId
  GROUP BY l1.License ),
Temp2(License2, Trajs) AS (
  SELECT l2.License, 
    collect_gs(list(trajectory_gs(t2.Trip)))
  FROM Trips t2, Licenses2 l2
  WHERE t2.VehicleId = l2.VehicleId
  GROUP BY l2.License )
SELECT License1, License2, 
  distance_gs(t1.Trajs, t2.Trajs) AS MinDist
FROM Temp1 t1, Temp2 t2  
ORDER BY License1, License2;
\end{lstlisting}

This version uses \texttt{trajectory\_gs()}, a version of \texttt{trajectory()} that returns a \texttt{GSERIALIZED} object as a \texttt{BLOB} in DuckDB instead of the well-known binary \texttt{WKB\_BLOB} format. \texttt{collect\_gs()}, a modified version of \texttt{ST\_Collect()}, then takes an array of these geometries and aggregates them into a collection. Finally, \texttt{distance\_gs()} takes two geometries, still in \texttt{GSERIALIZED} format, and returns the distance between them. This optimized query overcomes the drawback of the current interface with Spatial's \texttt{GEOMETRY} data type by utilizing MEOS' PostGIS-based functions.

\vspace{1mm}\noindent
\textbf{Query 7:} \textit{What are the license plate numbers of the passenger cars that have reached the points from \texttt{Points} first of all passenger cars during the complete observation period?}

\begin{lstlisting}[language=SQL,numbers=none]
WITH Timestamps AS (
  SELECT DISTINCT v.License, p.PointId, p.Geom,
    MIN(startTimestamp(atValues(t.Trip,
      p.Geom::WKB_BLOB))) AS Instant
  FROM Trips t, Vehicles v, Points1 p
  WHERE t.VehicleId = v.VehicleId AND
    v.VehicleType = 'passenger' AND
    t.Trip && stbox(p.Geom::WKB_BLOB) AND
    ST_Intersects(trajectory(t.Trip)::GEOMETRY, p.Geom)
  GROUP BY v.License, p.PointId, p.Geom )
SELECT t1.License, t1.PointId, t1.Geom, t1.Instant
FROM Timestamps t1
WHERE t1.Instant <= ALL (
  SELECT t2.Instant
  FROM Timestamps t2
  WHERE t1.PointId = t2.PointId )
ORDER BY t1.PointId, t1.License;
\end{lstlisting}

This query first creates the common table expression (CTE) \texttt{Timestamps} containing the vehicle information and the timestamps at which a passenger car reaches the points. In the \texttt{SELECT} statement, the query utilizes the \texttt{startTimestamp()} and \texttt{atValues()} functions. The \texttt{atValues()} function takes in a temporal value (in this case, temporal point geometry \texttt{tgeompoint} from \texttt{Trips}) and a base value (in this case, point geometry from \texttt{Points1}) to return the temporal value restricted to the second argument. Essentially, this function takes the full trip and returns only the temporal geometry values at the points from \texttt{Points1}. This value is then passed to \texttt{startTimestamp()} which, as the name suggests, returns the start timestamp of the temporal value, equivalent to the earliest timestamp at which a trip reaches any point from \texttt{Points1}. The third join condition utilizes the \textit{overlaps} (\texttt{\&\&}) predicate to filter trips that overlap with the point geometry by first creating a spatiotemporal bounding box (\texttt{stbox}) around the point.

\vspace{1mm}\noindent
\textbf{Query 10:} \textit{When and where did the vehicles with license plate numbers from \texttt{Licenses1} meet other vehicles (distance < 3 meters) and what are the latter licenses?}

\begin{lstlisting}[language=SQL,numbers=none]
WITH Temp AS (
  SELECT l1.License AS License1,
    t2.VehicleId AS Car2Id,
    whenTrue(tDwithin(t1.Trip, t2.Trip, 3.0)) AS Periods
  FROM Trips t1, Licenses1 l1, Trips t2, Vehicles v
  WHERE t1.VehicleId = l1.VehicleId AND
    t2.VehicleId = v.VehicleId AND
    t1.VehicleId <> t2.VehicleId AND
    t2.Trip && expandSpace(t1.trip::STBOX, 3.0) )
SELECT Licence1, Car2Id, Periods
FROM Temp
WHERE Periods IS NOT NULL;
\end{lstlisting}

This query first creates the CTE \texttt{Temp} to store the \texttt{Periods} when the vehicles met other vehicles within the spatial constraint. The \texttt{tDwithin()} function used in the projection is one of the spatial relationships generalized for temporal geometries. This function first computes, at each instant, whether the distance between the temporal points (\texttt{Trip} values from the \texttt{Trips} table) is less than or equal to 3. The function yields a \texttt{tbool} (temporal boolean) value representing the condition at all time instants of the trips. The resulting \texttt{tbool} value is passed to \texttt{whenTrue()}, which returns the time when the temporal boolean takes the value \texttt{true} as a \texttt{tstzspanset} value. This mobility type represents sets of ranges of \texttt{timestamptz} values. To filter out trips that are very far from each other, the \texttt{expandSpace()} function is used in the fourth join condition. The trip (from \texttt{t2}) is first cast into the \texttt{stbox} type, representing a spatiotemporal bounding box around the whole trip. \texttt{expandSpace()} expands the spatial dimension of this bounding box by 3 units. Only trips that overlap with this expanded box, filtered using the \textit{overlaps} (\texttt{\&\&}) predicate, are kept.

\subsubsection{Results and Discussions}

\begin{figure*}[h!]
    \centering
    \begin{subfigure}{0.47\textwidth}
        \includegraphics[width=\textwidth]{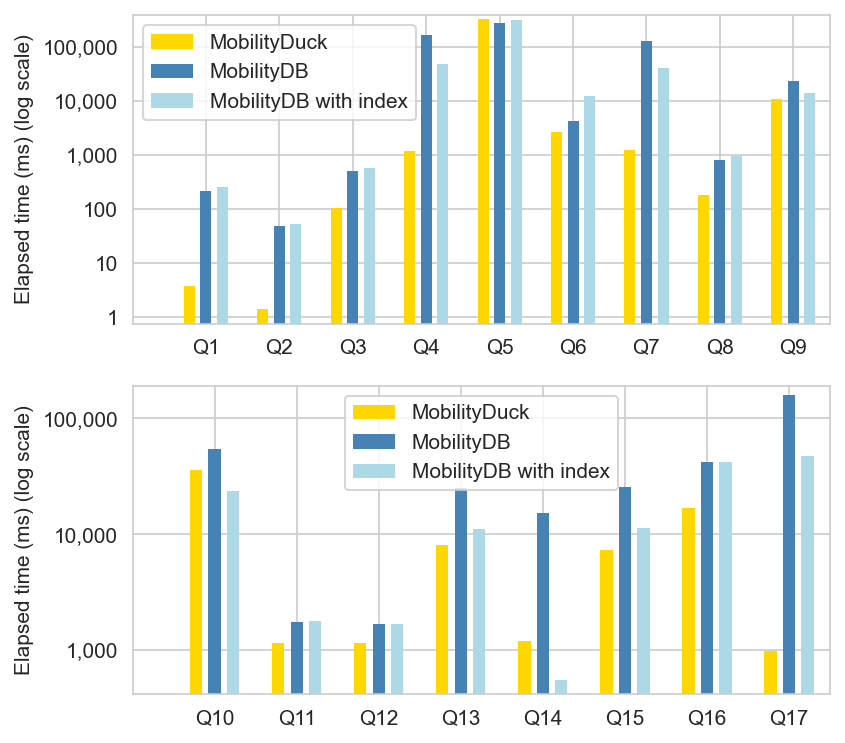}
        \caption{Query runtimes at SF-0.001}
        \hfill
    \end{subfigure}
    \begin{subfigure}{0.47\textwidth}
        \includegraphics[width=\textwidth]{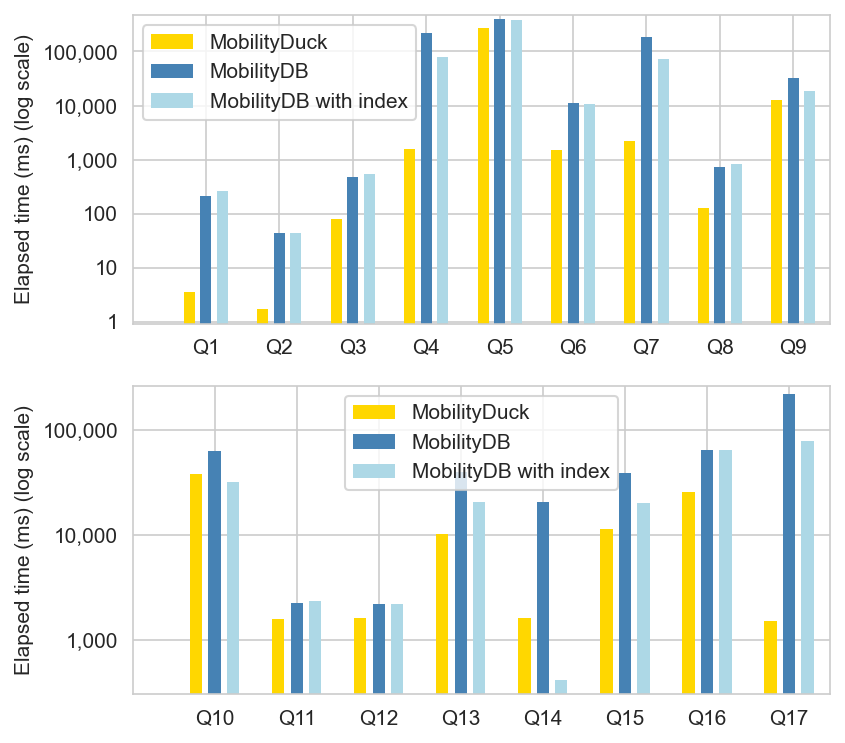}
        \caption{Query runtimes at SF-0.002}
        \hfill
    \end{subfigure}
    \begin{subfigure}{0.47\textwidth}
        \includegraphics[width=\textwidth]{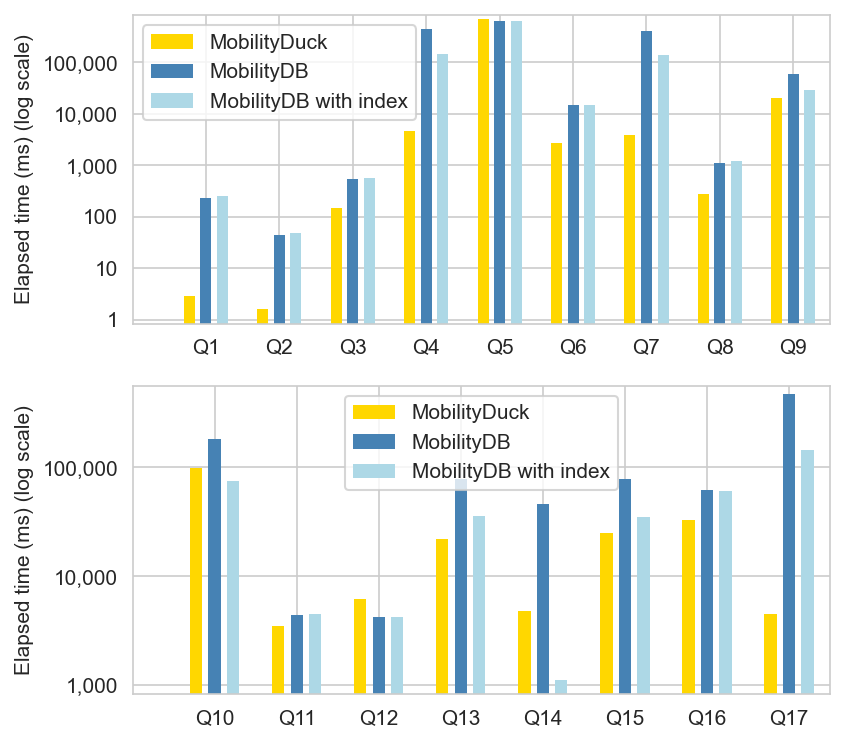}
        \caption{Query runtimes at SF-0.005}
    \end{subfigure}
    \begin{subfigure}{0.47\textwidth}
        \includegraphics[width=\textwidth]{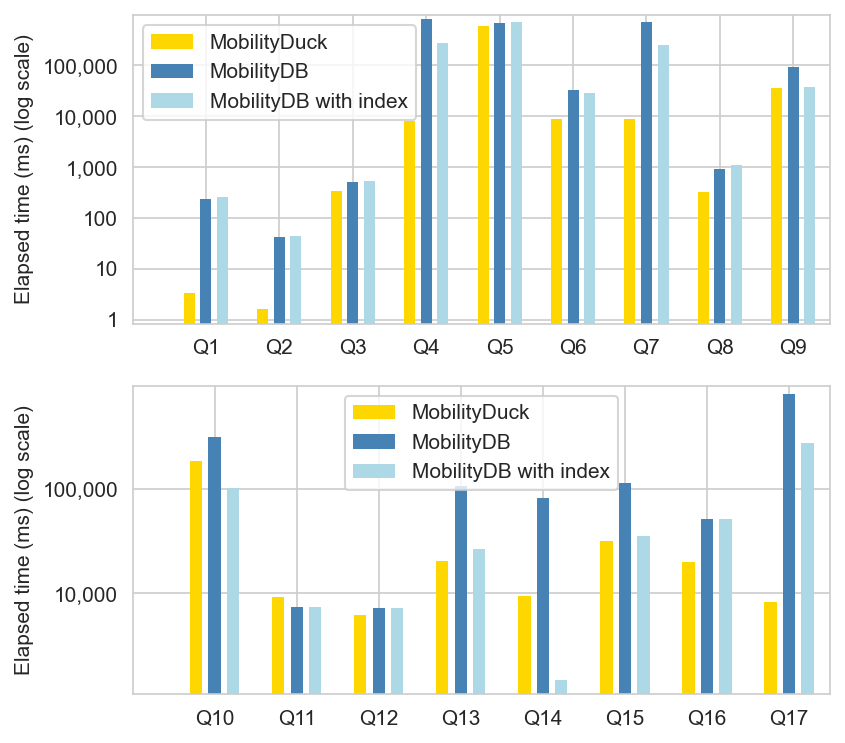}
        \caption{Query runtimes at SF-0.01}
    \end{subfigure}
    \caption{Runtimes in milliseconds for the BerlinMOD-Hanoi benchmark queries at SF-0.001, SF-0.002, SF-0.005, and SF-0.01}
    \label{fig:benchmark-results}
\end{figure*}

Figure~\ref{fig:benchmark-results} visualizes the runtimes of all 17 queries, across 4 scale factors, and for 3 scenarios: using MobilityDuck on DuckDB (yellow bars), using MobilityDB on PostgreSQL without indexes (dark blue bars), and using MobilityDB with indexes (light blue bars).

MobilityDuck outperforms MobilityDB both with and without indexes in all scale factors in 12 out of 17 queries (1, 2, 3, 4, 6, 7, 8, 9, 13, 15, 16, 17). For Query 5, MobilityDuck still manages to achieve the best runtimes in scale factors SF-0.002 and SF-0.01, though generally, the 3 scenarios achieve very similar runtimes for this query. For Query 10, across 4 scale factors, MobilityDuck is faster than MobilityDB without indexes, but slower than the case with indexes. The same applies for Query 14, where MobilityDuck, while being about 10 times faster than MobilityDB without indexes, is still slower than MobilityDB with indexes. For Query 11, MobilityDuck achieves the best runtimes in all scale factors except SF-0.01. Quite similarly, for Query 12, MobilityDuck is the best-performing system in all scale factors except SF-0.005. When compared to MobilityDB without indexes, MobilityDuck achieves higher performance in all cases except the following: Query 5 SF-0.001 (321.390 seconds versus 278.128 seconds) and SF-0.005 (682.170 seconds versus 620.794 seconds), Query 12 SF-0.005 (6.19 seconds versus 4.215 seconds), and Query 11 SF-0.01 (9.25 seconds versus 7.431 seconds).

Overall, in the majority of cases, MobilityDuck (plain, without indexes) outperforms MobilityDB, and in many cases, outperforms MobilityDB with indexes. These results demonstrate the effectiveness of integrating spatiotemporal querying directly within DuckDB's analytical engine, rather than relying on external index-based acceleration. The strong performance of MobilityDuck, even in the absence of specialized indexes, suggests an effective utilization of DuckDB's architecture to handle mobility workloads.


\section{Limitations and Future Work}
The latest implementation of MobilityDuck includes integrating MEOS and binding spatiotemporal types and functions adapted from the implementation of MobilityDB. So far, we have implemented many of the available types and functions in MEOS, but not all (see Table~\ref{tab:types}). As MEOS and MobilityDB are both constantly evolving with frequent additions of types and functionalities, the volume of such adaptation can grow very rapidly. Future development of MobilityDuck can benefit from an automated tool for generating bindings of all types and functions to ensure the most complete and up-to-date implementation on a par with MEOS.

In working with \texttt{GEOMETRY} type, we do not work with this type directly due to its specialized implementation by the Spatial extension. In the latest MobilityDuck implementation, we use an additional proxy layer where the functions that are supposed to return \texttt{GEOMETRY} type will, instead, return either \texttt{WKB\_BLOB} or \texttt{VARCHAR} types, which are standardized. The casting to and from these types and \texttt{GEOMETRY} is left for the Spatial extension to handle, which is enforced by adding \texttt{::GEOMETRY}, \texttt{::WKB\_BLOB}, etc. to the relevant values. This interface, while simple in terms of implementation, results in unnecessary overheads when dealing with data of type \texttt{GEOMETRY}, as previously discussed in Section~\ref{benchmark-results}. Future development of MobilityDuck will focus on a more refined integration with the Spatial extension in order to support the \texttt{GEOMETRY}-related functions more natively and efficiently.

MobilityDuck currently does not support the \texttt{geography} type. Future work will examine the native support for this data type, such as with the use of the Geography extension,\footnote{\url{https://duckdb.org/community_extensions/extensions/geography.html}} in order to develop MobilityDuck's interface for handling this type.

\section{Conclusion}
This paper introduces MobilityDuck, a DuckDB extension which integrates the mobility data management capacity of MEOS into a lightweight, in-memory analytical database system. By embedding spatiotemporal types and trajectory operators directly into an in-memory analytical engine, MobilityDuck is among the first systems to bridge the gap between traditional moving object databases and modern embedded analytics systems. It enables efficient analysis of large spatiotemporal datasets while preserving DuckDB’s strengths in performance, simplicity, and seamless integration with data science environments. By using the BerlinMOD-Hanoi dataset, we demonstrated the performance of MobilityDuck. First, we presented a use case scenario that integrates MobilityDuck with DuckDB's Python API and the Jupyter Notebook environment, enabling fast visualizations of query results and showing the quick integration of MobilityDuck into the existing DuckDB ecosystem. Secondly, we compared the runtime of MobilityDuck against MobilityDB using a set of benchmark queries of BerlinMOD. In the majority of cases, MobilityDuck achieved better results than MobilityDB both with and without indexes, showing its potential for developing a unified, high-performance analytical framework for spatiotemporal data.

For future work, we aim to expand the spatiotemporal analytics capabilities of MobilityDuck by adding support for the remaining types and functions of MEOS, and potentially develop an automated tool for keeping MobilityDuck up-to-date with both MEOS and MobilityDB. Additionally, we plan to further develop the indexing capabilities of MobilityDuck to support indexing more spatiotemporal data types, as well as to re-evaluate its performance once indexing has been more thoroughly supported.

\section{Artifacts}\label{artifacts}
All relevant code and instructions for building MobilityDuck as long as the use case demonstration and benchmark are available on the official GitHub repository.\footnote{\url{https://github.com/MobilityDB/MobilityDuck}} In addition, precompiled binary extension packages are also provided in GitHub repository for Linux, macOS, and DuckDB-Wasm. At present, Windows is not supported due to compatibility limitations in the underlying MEOS library. Support for Windows will be made available once MEOS provides the necessary compatibility.
All code and data related to BerlinMOD-Hanoi are also available.\footnote{\url{https://github.com/MobilityDB/MobilityDB-BerlinMOD-Hanoi}}

\bibliographystyle{ACM-Reference-Format}
\bibliography{references}
\end{document}